\documentclass[journal]{IEEEtran}
\usepackage{physics}
\usepackage{graphicx}
\usepackage{float}
\usepackage{amsmath}
\usepackage{subfigure}
\usepackage{amsmath}
\usepackage{longtable}
\usepackage{booktabs}
\usepackage{float}
\usepackage{threeparttable}
\usepackage{xcolor}
\usepackage{diagbox}
\usepackage{cite}
\usepackage{hyperref}

\ifCLASSINFOpdf
\else
\fi

\begin{document}
%
\title{FAVER: Blind Quality Prediction of \\Variable Frame Rate Videos}
%
%
%

\author{Qi~Zheng$^{\dagger}$,
        Zhengzhong~Tu$^{\dagger}$,
        Pavan~C.~Madhusudana,
        Xiaoyang~Zeng, Alan~C.~Bovik,~\IEEEmembership{Fellow,~IEEE,
        Yibo~Fan$^\star$}
\thanks{Qi Zheng, Yibo Fan and Xiaoyang Zeng are with the State Key Laboratory of ASIC \& System, College of Microelectronics, Fudan University, Shanghai 200000, China (e-mail: qzheng21@m.fudan.edu.cn; fanyibo@fudan.edu.cn; xyzeng@fudan.edu.cn).}
\thanks{Zhengzhong Tu, Pavan C. Madhusudana, and Alan C. Bovik are with the Laboratory for Image and Video Engineering (LIVE), Department of Electrical and Computer Engineering, The University of Texas at Austin, Austin, TX 78712 USA (e-mail: zhengzhong.tu@utexas.edu; bovik@utexas.edu).}
\thanks{$^\star$ Corresponding author: Yibo Fan (e-mail: fanyibo@fudan.edu.cn).}
\thanks{$^{\dagger}$ Joint first authors.}
}

%
%

\markboth{Journal of \LaTeX\ Class Files,~Vol.~14, No.~8, August~2015}%
{Shell \MakeLowercase{\textit{et al.}}: Bare Demo of IEEEtran.cls for IEEE Journals}
%



\maketitle 
\begin{abstract}
Video quality assessment (VQA) remains an important and challenging problem that affects many applications at the widest scales. Recent advances in mobile devices and cloud computing techniques have made it possible to capture, process, and share high resolution, high frame rate (HFR) videos across the Internet nearly instantaneously. Being able to monitor and control the quality of these streamed videos can enable the delivery of more enjoyable content and perceptually optimized rate control. Accordingly, there is a pressing need to develop VQA models that can be deployed at enormous scales. While some recent effects have been applied to full-reference (FR) analysis of variable frame rate and HFR video quality, the development of no-reference (NR) VQA algorithms targeting frame rate variations has been little studied. Here, we propose a first-of-a-kind blind VQA model for evaluating HFR videos, which we dub the Framerate-Aware Video Evaluator w/o Reference (FAVER). FAVER uses extended models of spatial natural scene statistics that encompass space-time wavelet-decomposed video signals, to conduct efficient frame rate sensitive quality prediction. Our extensive experiments on several HFR video quality datasets show that FAVER outperforms other blind VQA algorithms at a reasonable computational cost. To facilitate reproducible research and public evaluation, an implementation of FAVER is being made freely available online: \url{https://github.com/uniqzheng/HFR-BVQA}. 
\end{abstract}



%
\IEEEpeerreviewmaketitle

\section{Introduction}
\label{sec:intro}
%
%
%
%
\IEEEPARstart{B}{ecause} of the rapid development of streaming media technologies coupled with the increasing popularity of social media platforms like YouTube, Instagram, and TikTok, videos now play a central role in the daily lives of billions of people. One of the technological enablers of this sea change are video compression techniques that are able to optimize tradeoffs of high perceptual quality against efficient bitrate use. Perceptual optimization of streaming media remains a fundamental research topic that continue to lead to models and algorithms significantly impact workflows in the multimedia industry. Video quality assessment (VQA) tools are widely used to monitor and control video encoders, communication systems, and enhancement and reconstruction algorithms. VQA algorithms are the essential tools for monitoring and optimizing the visual quality-of-experience (QoE) of billions of streaming and social media customers.  

Depending on the amount of reference information required by the model, video quality assessment can be roughly classified into two categories: reference (R) and no-reference (NR) video quality models. R-VQA models compute perceptual differences between reference videos against their distorted counterparts. These methods include such widely deployed models as peak signal to noise ratio (PSNR), SSIM~\cite{wang2004image}, MOVIE ~\cite{seshadrinathan2009motion}, and VMAF~\cite{Vmaf}. In practice, however, it is sometimes impossible to have access to any reference signal. This is generally the case when designing video quality assessment models for user-generated content (UGC-VQA)~\cite{9405420}, which may suffer from complexes of in-capture and/or post-capture distortions. In such scenarios, no-reference or blind VQA models (NR-VQA/BVQA) are the required tools, although reference VQA models may be used in concert with NR VQA algorithms if distorted UGC content is to be compressed under perceptual control~\cite{yu2021predicting}.

The most successful and widely used general-purpose NR-VQA/IQA models~\cite{mittal2012no,6353522,FRIQUEE} are based on modifications of natural scene statistics (NSS) models~\cite{ruderman1994statistics} of bandpass luminance~\cite{mittal2012no,6353522,FRIQUEE,ruderman1994statistics,xue2014blind} and chrominance signals ~\cite{FRIQUEE, kundu2017no, 9405420, tu2021rapique}. More recently, more sophisticated temporal natural video statistics models of bandpass (displaced) frame-difference signals~\cite{lee2021space, 9524440}, or temporal filter responses~\cite{saad2014blind, mittal2015completely, korhonen2019two, 9405420} have also been explored. Convolutional neural networks (CNNs or ConvNets) have been shown to deliver promising results on video quality assessment problems, because of the availability of modern large-scale databases~\cite{hosu2017konstanz, sinno2018large, wang2019youtube, ying2020patches, li2019quality,Ying_2021_CVPR}. Several recent BVQA algorithms based on pre-trained semantic ConvNet features fine-tuned on large subjective databases have been leveraged to extract both content- and distortion-aware spatio-temporal features, resulting in state-of-the-art quality prediction performance~\cite{li2019quality,Ying_2021_CVPR,Wang_2021_CVPR,liu2018end}.

A topic of very recent interest is the possibility of augmenting adaptive bitrate (ABR) streaming by allowing control over variable frame rates (VFR). This concept has increased relevance given the availability of high-frame rate (HFR) videos which can improve viewing experiences on high-motion and rapidly changing videos. For example, just as streaming providers now routinely alter spatial resolution along with standard compression in the construction of bitrate ladders, they could also modify the streamed frame rate before compression depending on the content characteristics. Creating ABR bitrate ladders that allow for VFRs will require new VQA models and algorithms to perform perceptual optimization. In this direction, new databases have been built containing VFR contents~\cite{mackin2018study,madhusudana2021subjective}, and efficient predictors of compressed VFR video quality have been created~\cite{lee2021space,9524440}.

Just as there is now a heightened need for FR VQA models for VFR/HFR scenarios, so also is there for blind/NR models that can address the effects of frame rate changes throughout video workflows without the availability of any reference. Yet, existing BVQA models have been focused on fixed frame-rate videos, and there are no blind models specifically designed to predict the quality of VFR/HFR videos. Although a variety of temporal ``quality-aware'' features have been explored~\cite{lee2021space, 9405420, tu2021rapique, Ying_2021_CVPR}, these are generally between-frame measurements with no demonstrated efficacy on VFR videos. Since the demand for HFR videos is increasing and many soon explode, so also will VFR protocols and the need for video quality prediction models capable of guiding them, including those that lead to BVQA algorithms.

Here we propose, to the best of our knowledge, \textit{the first NR-VQA} model that targets visual distortions arising from framerate variations, which we dub the Framerate Aware Video Evaluator w/o Reference, or FAVER. The design of FAVER relies on a general, effective and efficient temporal statistics model that is based on the well-established bandpass regularities of natural videos. Our experiments show that FAVER achieves state-of-the-art performance on predicting frame-rate-variant video quality on VFR/HFR databases.

The rest of this paper is organized as follows. In Section~\ref{sec:related-work}, we provide a discussion of background and related work on video quality assessment. Section~\ref{sec:proposed-method} describes the details of our proposed FAVER model. Experimental settings and results are presented Section~\ref{sec:experimental}, and Section~\ref{sec:conclusion} concludes the paper with some final remarks.

\section{Related Work}
\label{sec:related-work}

\subsection{Subjective Video Quality Assessment}
\label{ssec:subjective-vqa}

There has been extensive research on subjective video quality assessment over the past two decades. Early subjective video quality studies like the LIVE-VQA~\cite{seshadrinathan2010study}, CSIQ-VQA~\cite{vu2014vis3}, and VQEG database~\cite{vqeg_hdtv} were built to target specific types of common single distortions, such as compression and transmission errors. However, real-world videos potentially contain multiple unknown but commingled distortions, including but not limited to compression, camera motion, and a great variety of capture defects, which are not at all represented by these legacy VQA datasets. CVD2014\cite{7464299} was perhaps the first VQA dataset to contain authentic natural scene distortions of realistic contents captured by a variety of devices. More recently, the Konstanz (KoNViD-1k)~\cite{7965673}, LIVE-Qualcomm~\cite{7932975}, LIVE Video Quality Challenge (LIVE-VQC)\cite{8463581}, YouTube-UGC~\cite{wang2019youtube}, and LIVE-Facebook~\cite{Ying_2021_CVPR} datasets have greatly expanded the volume and variety of distorted contents, allowing for the development of more general NR video quality assessment engines suitable for user-generated content (UGC)~\cite{9405420}. 

Datasets designed for analyzing HFR contents have been relatively limited. An early HFR video quality database (Waterloo-IVC HFR)~\cite{7340831} contains 480p and 1080p videos with a variety of frame rates ranging from 5 to 60, distorted by four compression levels using H.264. Unfortunately, this database is not publicly available. The authors of BVI-HFR~\cite{8531714} introduced a database consisting of 22 video contents having frame rates of 15, 30, 60, and 120. Madhusudana et al. proposed the LIVE-YT-HFR~\cite{madhusudana2021subjective} database which encompasses 16 unique contents distorted by a variety of combinations of frame rates (24, 30, 60, 82, 98, 120 fps) and several VP9 compression levels, yielding a total of 480 test videos. This database also has the advantage of including videos of both popular resolutions (1080p and 4K), allowing for the creation and testing of VFR VQA models that are able to handle both popular streaming resolutions. Note that BVI-HFR and LIVE-YT-HFR differ in their temporal subsampling schemes: the authors of BVI-HFR obtained reduced frame rate videos using frame averaging, while the LIVE-YT-HFR authors adopted frame-dropping, resulting in slightly different motion artifacts.

\subsection{Objective BVQA Models}
\label{ssec:objective-vqa}
\subsubsection{Feature-Based BVQA Models}
\label{sssec:traditional-bvqa}

The earliest BVQA / BIQA algorithms were designed to analyze and quantify a single distortion type, such as blockiness, blur, ringing, banding, and compression~\cite{5246972, 6476013, wang2002no, wang2000blind, tu2020bband}, based on the measurement of a small number of image/frame level features. For example, CPBDM~\cite{5246972} and LPCM~\cite{6476013} were developed for blur evaluation, while NJQA~\cite{6691937} and JPEG-NR~\cite{wang2002no} aimed to assess noise and JPEG compression, respectively. These have been largely replaced by much more powerful general-purpose BIQA/BVQA models that are based on measurements of distortion-induced deviations of bandpass processed images/videos from perceptually relevant natural scene statistics (NSS)~\cite{ruderman1994statistics}. These approaches succeed because high-quality naturalistic images and videos reliably exhibit certain statistical regularities, which are predictably altered by distortions~\cite{mittal2015completely, mittal2012no}. For example, a small set of (36) spatial bandpass and locally normalized spatial features are computed by BRISQUE~\cite{mittal2012no}, which uses a support vector regressor (SVR) to map them to human subjective judgements of quality. NIQE~\cite{6353522} maps the same 36 features without a learning process, by computing a simple statistical distance from typical (high quality) values of these features. Similarily, GM-LOG~\cite{xue2014blind} extracts statistical features in a smoothed gradient and Laplacian-of-Gaussian domain, while HIGRADE~\cite{kundu2017no} extracts gradient features in an LAB color space. A larger variety of color spaces and perception-driven transforms were utilized to extract a larger number perceptually relevant NSS features in a ``bag of features'' model called FRIQUEE~\cite{FRIQUEE}. 

\begin{figure*}[!t]
\centering
\footnotesize
\def\xheight{0.75}
\setlength{\tabcolsep}{1pt}
    \includegraphics[width=\xheight\linewidth]{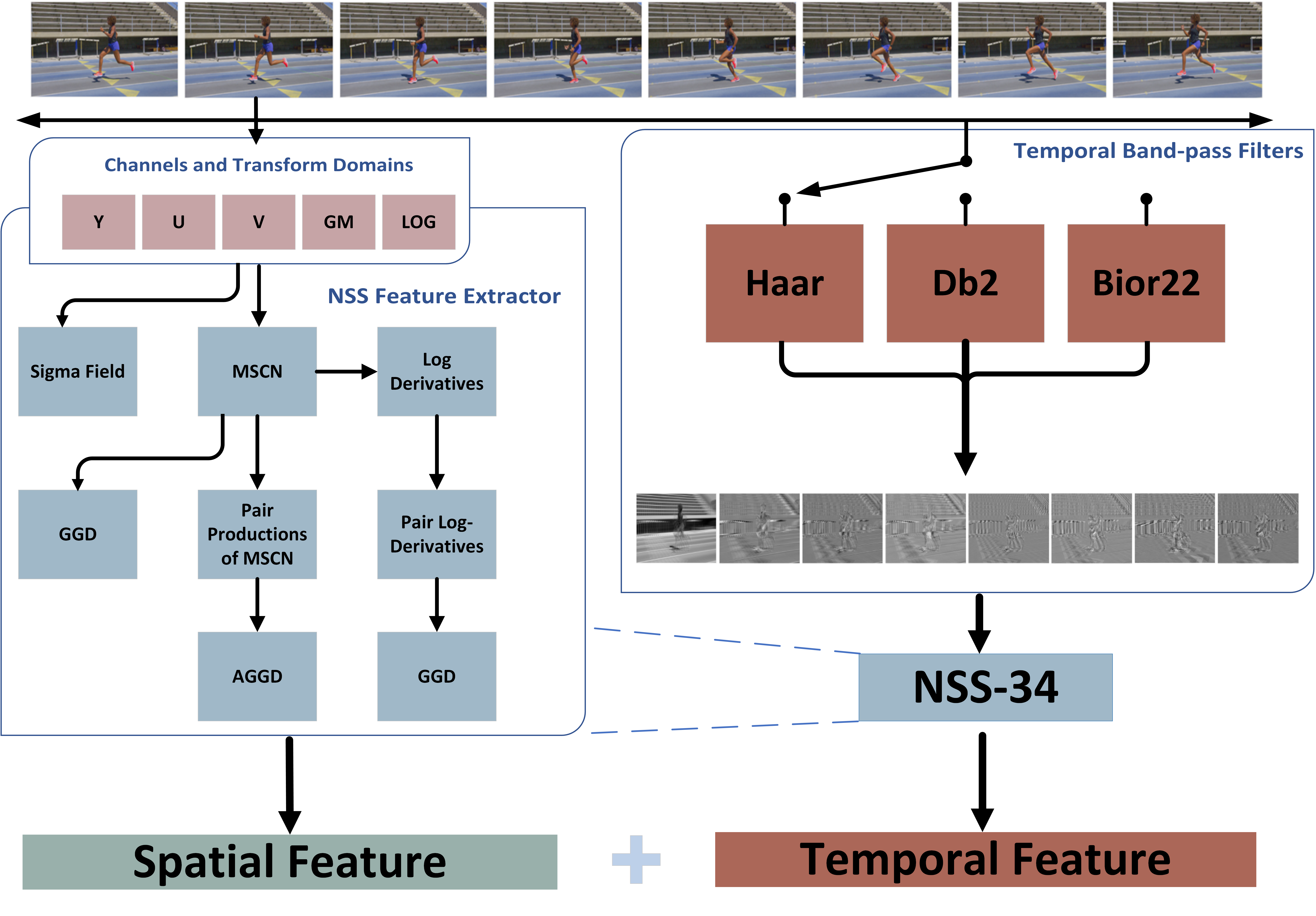}
\caption{The overall framework of the proposed FAVER model. The left block shows the spatial feature extraction path while the right block depicts the temporal feature extraction branch. A shared  NSS feature extraction model processes both the spatial and temporal feature responses.}
\label{fig:framework}
\end{figure*}

Blind VQA models have also been devised that make measurements of the temporal statistics of natural videos. V-BLIINDS\cite{6705673} utilizes a spatio-temporal statistics model of DCT coefficients of frame-differences to predict perceptual video quality scores. Li et al.~\cite{li2016spatiotemporal} successfully compute spatio-temporal natural video statistics in the 3D-DCT domain. TLVQM~\cite{korhonen2019two} computes handcrafted quality-aware features including motion-based statistics and aesthetics features to train a regressor to predict a video quality score. Inspired by the success of VMAF~\cite{Vmaf}, VIDEVAL~\cite{9405420} uses a similar ensemble model that fuses features taken from several successful BVQA models, based on a feature selection procedure.

\subsubsection{Deep Learning-Based BVQA Models}
\label{sssec:deep-learning-bvqa}

Deep convolutional neural networks (CNNs or ConvNets) have been shown to deliver superior performance on numerous image analysis tasks~\cite{chen2020proxiqa, jiang2021enlightengan, kupyn2019deblurgan}. Recently, several successful deep learning-based BVQA models have emerged. VSFA~\cite{li2019quality} seeks to measure content dependencies and temporal-memory to learn an end-to-end deep neural network NR-VQA model. V-MEON~\cite{liu2018end} merges feature extraction and regression into a single stage where both are jointly optimized within a multi-task DNN. Among hybrid models, RAPIQUE~\cite{tu2021rapique} efficiently combines low-level NSS features with high-level deep learning features which are used to train a single regressor head that predicts video quality scores. PVQ~\cite{Ying_2021_CVPR} extracts both spatial and temporal features using RoIPool (region-of interest pooling) and SoIPool (segment-of-interest pooling) layers, using them to learn local to global spatio-temporal quality relationships. CoINVQ~\cite{Wang_2021_CVPR} decomposes UGC quality understanding into several sub-tasks that are learned using three subnets (called ContentNet, DistortionNet, and CompressionNet) that extract 2D deep features to make quality predictions. 

\section{Proposed Method}
\label{sec:proposed-method}

While there are now available a variety of BVQA models that are able to capture various temporal aspects of perceptual video quality, none have been designed to handle VFR/HFR video quality measurement. Moreover, in most of these prior models, the composition of temporal features largely extends spatial features to the time domain, e.g., by applying them on frame differences. There is evidence that simple between-frame measurements are inadequate to capture frame rate induced visual distortions~\cite{9524440}. Therefore, it is of great interest to consider the design of no-reference VQA models applicable to videos capable of capturing temporal aspects of perceptual quality arising from frame rate variations.

Towards advancing progress in this direction, we have designed a VFR-sensitive BVQA model (FAVER) whose processing flow is depicted in Fig.~\ref{fig:framework}. FAVER is defined by two branches that respectively contain spatial feature extraction and temporal feature calculation modules. The latter module comprises temporal bandpass filters, the responses of which are expressive of the expected statistical regularities of videos, which can be affected by frame rate. Next, we describe the spatial and temporal branches of FAVER.

\subsection{Spatial Feature Design}
\label{ssec:Spatial-feature}
The spatial features that constitute the spatial quality evaluation component of FAVER have all been previously used in other models~\cite{mittal2012no,6353522,FRIQUEE},~\cite{xue2014blind,kundu2017no,DBLP}. While they contribute strongly to the efficacy of FAVER, our main contributions are the temporal features described further along, and how they are integrated with the spatial features.
\subsubsection{Feature Extraction Module}
\label{sssec:feature-extraction-module} 

It has been shown that the Mean Subtracted and Contrast Normalized (MSCN) coefficients of images or videos present empirical distributions that are sensitive to visual perceptual quality~\cite{mittal2012no,6353522,FRIQUEE, kundu2017no, 7094273, DBLP}. Specifically, consider given luminance image $I(i,j)$, the MSCN coefficients are
\begin{equation}
\label{eq:mscn}
    \widehat{I}(i,j) = \frac{I(i,j) - \mu(i,j)}{\sigma(i,j) + C},
\end{equation}
where $i \in 1,2,...M$, $j \in 1,2...N$, are spatial indices, and $C=1$ is a saturation constant that reduces instabilities. In (\ref{eq:mscn}),  $\mu(i,j)$ and $\sigma(i,j)$ are local weighted sample means and standard deviations given by 
\begin{equation}
    \mu(i,j) = \sum_{k=-K}^K \sum_{l=-L}^L \omega_{k,l} I_{k,l}(i,j),
\end{equation}
and
\begin{equation}
    \sigma(i,j) = \sqrt{\sum_{k=-K}^K \sum_{l=-L}^L \omega_{k,l} (I_{k,l} (i,j) - \mu(i,j) )^2},
\end{equation}
where $\omega={\omega_{k,l}|k = -K,...,K, l = -L,...,L}$ is a 2D
circularly-symmetric Gaussian weighting function sampled out to 3 standard deviations and rescaled to unit volume. In our implementation, $K = L = 3$.

The first-order statistics of the MSCN coefficients of high quality natural images/videos strongly tend towards a Gaussianity. Visual distortions/degradations generally alter this statistical regularity in ways that can be used to accurately predict perceived quality~\cite{mittal2012no, 6353522}. This simple yet effective observation has led to a wide variety of IQA/VQA models that map measured statistical model parameters to perceptual quality in different bandpass domains~\cite{mittal2012no,6353522,FRIQUEE,xue2014blind,kundu2017no}. One very recent model called RAPIQUE~\cite{tu2021rapique} adopts a particularly efficient strategy, whereby instead of hand-crafting features in a variety of bandpass spaces, it deploys an NSS-based feature extractor which is applied on each feature transform. RAPIQUE's powerful and compute-efficient strategy attains SOTA performance across a wide variety of subjective databases. Inspired by the efficacy and modularity of RAPIQUE, we leverage the basic NSS feature extraction module used in RAPIQUE, applying it on several spatial feature maps to extract rich quality-aware statistical features.

The basic feature extractor is based on simple natural image statistics models. The first basic model is the zero-mean generalized Gaussian distribution (GGD):
\begin{equation}
    f(x;a,\sigma^2) = \frac{\alpha}{2\beta \varGamma(1/\alpha)}\exp(-(\frac{\left\lvert x \right\rvert}{\beta})^\alpha),
\end{equation}
\begin{equation}
    \beta = \sigma \sqrt{\frac{\varGamma(1/\alpha)}{\varGamma(3/\alpha)}},
\end{equation}
where the model parameters $\alpha$ and  $\sigma$ control shape and variance respectively, and $\varGamma(\cdot)$ is the gamma function. Models like BRISQUE and NIQE conduct quality prediction by finding the model parameters yielding the best GGD fits to empirical MSCN distributions.

Likewise, a statistical bivariate model relating neighboring pixels is obtained by constructing empirical distributions of pairwise products of adjacent MSCN coefficients along the horizontal, vertical, main diagonal, and secondary diagonal: ($H(i,j) = \widehat{I}(i,j)\widehat{I}(i,j+1)$, $V(i,j) = \widehat{I}(i,j)\widehat{I}(i+1,j)$, $ D1(i,j) = \widehat{I}(i,j)\widehat{I}(i+1,j+1)$, $D2(i,j) = \widehat{I}(i,j)\widehat{I}(i+1,j-1)$). Four additional quality-sensitive parameters are gathered by finding the best fitting asymmetric Generalized Gaussian distribution (AGGD) to the product histograms:
\begin{equation}
\label{eq:aggd}
f(x;\!\nu,\!\sigma_l^2,\!\sigma_r^2)\!=\!
\left\{\!
\begin{array}{ll}
\!\cfrac{\nu}{(\beta_l\!+\!\beta_r)\Gamma(\frac{1}{\nu})}\!\exp\!{\left(\!-\!\left(\!\cfrac{\!-\!x\!}{\beta_l}\right)^{\!\nu}\!\right)}\! & x\!<\!0 
\\[2ex]
\!\cfrac{\nu}{(\beta_l\!+\!\beta_r)\Gamma(\frac{1}{\nu})}\!\exp\!{\left(\!-\!\left(\!\cfrac{x}{\beta_r}\right)^{\!\nu}\right)}\! & x\!>\!0,
\end{array}
\right.
\end{equation}
where $\beta_l = \sigma_l \sqrt{\Gamma(1/\nu)/\Gamma(3/\nu)}$, and $\beta_r = \sigma_r \sqrt{\Gamma(3/\nu)/\Gamma(1/\nu)}$. The AGGD model has four parameters: $\nu$ controls the shape of the distribution while $(\sigma_l,\sigma_r)$ are scale parameters controlling the spread to the left and right of the origin respectively, and $\eta$  the mean of the distribution is by $\eta=(\beta_r-\beta_l)\Gamma(2/\eta)/\Gamma(1/\eta)$.

It is also valuable to compute the logarithmic derivative statistics of adjacent MSCN coefficient pairs~\cite{kundu2017no}, from which we also extract quality prediction features. Let 
\begin{equation}
    J(i,j) = \log[\widehat{I}(i,j) + C],
\end{equation}
where constant $C$ is introduced to prevent numerical instabilities. The gradient components of the log MSCN values along 7 orientations are calculated as:
\begin{equation}
\label{eq:log-der}
\begin{gathered}
    D1: \bigtriangledown _x J(i,j) = J(i,j+1) - J(i,j) \\
    D2: \bigtriangledown _y J(i,j) = J(i+1,j) - J(i,j) \\
    D3: \bigtriangledown _{xy} J(i,j) = J(i+1,j+1) - J(i,j) \\
    D4: \bigtriangledown _{yx} J(i,j) = J(i+1,j-1) - J(i,j) \\
    D5: \bigtriangledown _x \bigtriangledown _y J(i,j) = J(i-1,j) + J(i+1,j) \\
    - J(i,j-1) - J(i,j+1) \\
    D6: \bigtriangledown _{cx} \bigtriangledown _{cy} J(i,j)_1 = J(i,j) + J(i+1, j+1) \\
    - J(i,j+1) - J(i+1, j) \\
    D7: \bigtriangledown _{cx} \bigtriangledown _{cy} J(i,j)_2 = J(i-1,j-1) + J(i+1,j+1) \\
    - J(i-1,j+1) - J(i+1,j-1). \\
\end{gathered}
\end{equation}
The empirical distributions of each of these is then fitted with the best GGD model, yielding 14 additional shape and variance features.
\par
Returning to the original MSCN image (or frame) map (\ref{eq:mscn}), the ``$\sigma$-field'' $\sigma(i,j)$ has previously been shown to be a rich source of quality-related NSS features\cite{kundu2017no}\cite{FRIQUEE}. We extract two parameters from $\sigma$-field: the mean value $\phi_\sigma$ 
\begin{equation}
    \phi_\sigma = \frac{1}{MN} \sum_{i=0}^{M-1} \sum_{j=0}^{N-1} \sigma(i,j),
\end{equation}
and the square of the reciprocal of the coefficient of variation
\begin{equation}
    \rho_{\sigma}=(\phi_{\sigma}/\omega_{\sigma})^2,
\end{equation}
where 
\begin{equation}
\label{eq:omega}
    \omega_{\sigma}=\sqrt{\frac{1}{MN}\sum_{i=0}^{M-1}\sum_{j=0}^{N-1}[\sigma(i,j)-\phi_\sigma]^2}.
\end{equation}

The processing flow of the basic feature extraction module is presented in Table~\ref{table:feature-extraction}. Given the 34-dimensional NSS feature extraction module, feature subsets can be extracted on versions of the image/frame that have been processing spatially or temporally to arrive at a set of statistical features that account for various aspects of quality perception. 

\begin{table}
\caption{Summary of the basic 34-dimensional NSS feature extraction module.}
\label{table:feature-extraction}
\begin{tabular}{lrr}
\toprule
\textbf{Index}           & \textbf{Description}                    & \textbf{Computation Procedure}                 \\\midrule
$f_1-f_2$       & $(\alpha,\sigma)$              & Fit GGD to MSCN coefficients          \\
$f_3-f_4$       & $(\phi_\sigma,\rho_\sigma)$  & Compute statistics on `sigma' map     \\
$f_5-f_8$       & $(\nu,\eta,\sigma_l,\sigma_r)$ & Fit AGGD to H pairwise products       \\
$f_9-f_{12}$    & $(\nu,\eta,\sigma_l,\sigma_r)$ & Fit AGGD to V pairwise products       \\
$f_{13}-f_{16}$ & $(\nu,\eta,\sigma_l,\sigma_r)$ & Fit AGGD to D1 pairwise products      \\
$f_{17}-f_{20}$ & $(\nu,\eta,\sigma_l,\sigma_r)$ & Fit AGGD to D2 pairwise products      \\
$f_{21}-f_{22}$ & $(\alpha,\sigma)$              & Fit GGD to D1 pairwise log-derivative \\
$f_{23}-f_{24}$ & $(\alpha,\sigma)$              & Fit GGD to D2 pairwise log-derivative \\
$f_{25}-f_{26}$ & $(\alpha,\sigma)$              & Fit GGD to D3 pairwise log-derivative \\
$f_{27}-f_{28}$ & $(\alpha,\sigma)$              & Fit GGD to D4 pairwise log-derivative \\
$f_{29}-f_{30}$ & $(\alpha,\sigma)$              & Fit GGD to D5 pairwise log-derivative \\
$f_{31}-f_{32}$ & $(\alpha,\sigma)$              & Fit GGD to D6 pairwise log-derivative \\
$f_{33}-f_{34}$ & $(\alpha,\sigma)$              & Fit GGD to D7 pairwise log-derivative \\
\bottomrule
\end{tabular}
\end{table}

\begin{figure}[!t]
\centering
\footnotesize
\def\xheight{0.485}
\setlength{\tabcolsep}{1pt}
\begin{tabular}{cc}
{\includegraphics[width=\xheight\linewidth]{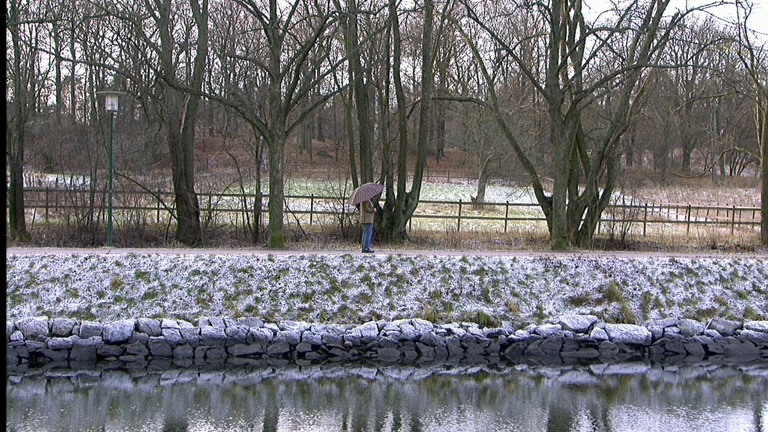}}  &
{\includegraphics[width=\xheight\linewidth]{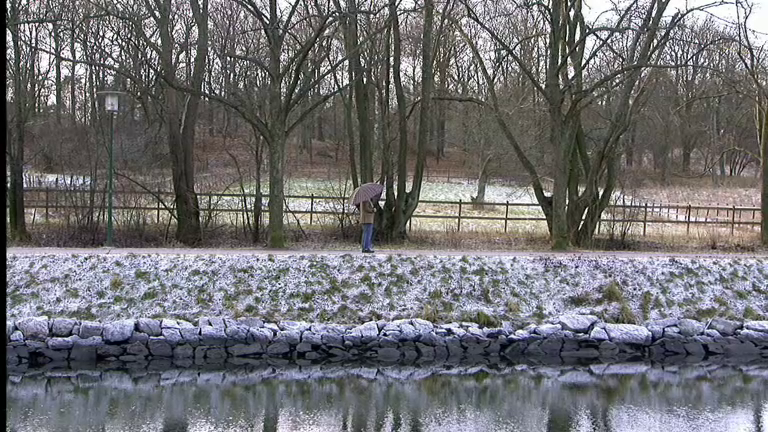}} \\
\texttt{PR01} Reference (DMOS = 0) & \texttt{PR09} Excellent quality (DMOS = 39.8) \\
{\includegraphics[width=\xheight\linewidth]{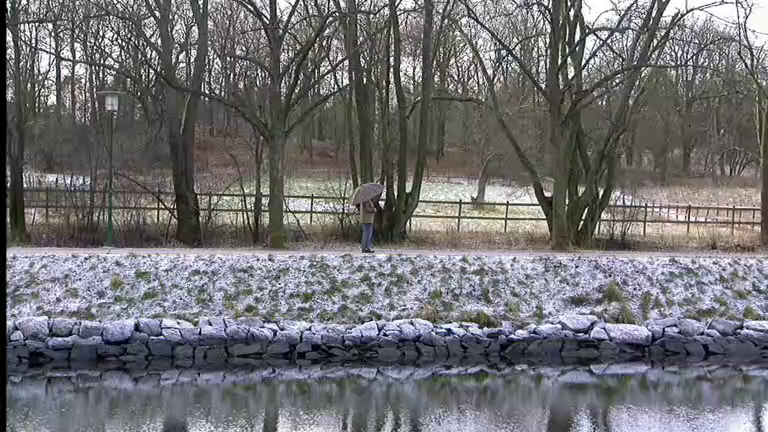}} &
{\includegraphics[width=\xheight\linewidth]{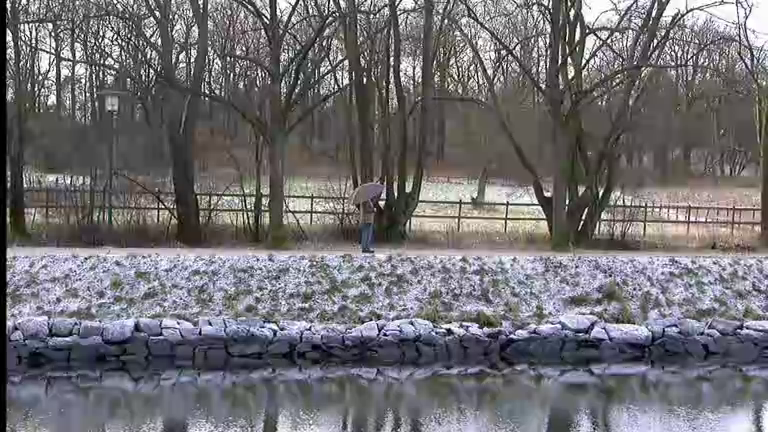}} \\
\texttt{PR11} Fair quality (DMOS = 59.9) & \texttt{PR12} Poor quality (DMOS = 77.3) \\ 
\end{tabular}
\caption{Four frames extracted at the 17th second from four H.264 compressed versions of the sequence \texttt{Park Run} from the LIVE-VQA~\cite{5404314} dataset}
\label{fig:sample-frame}
\end{figure}

\begin{figure}[!t]
\centering
\footnotesize
\def\xheight{0.55}
\setlength{\tabcolsep}{1pt}
\begin{tabular}{c}
\includegraphics[width=\xheight\linewidth]{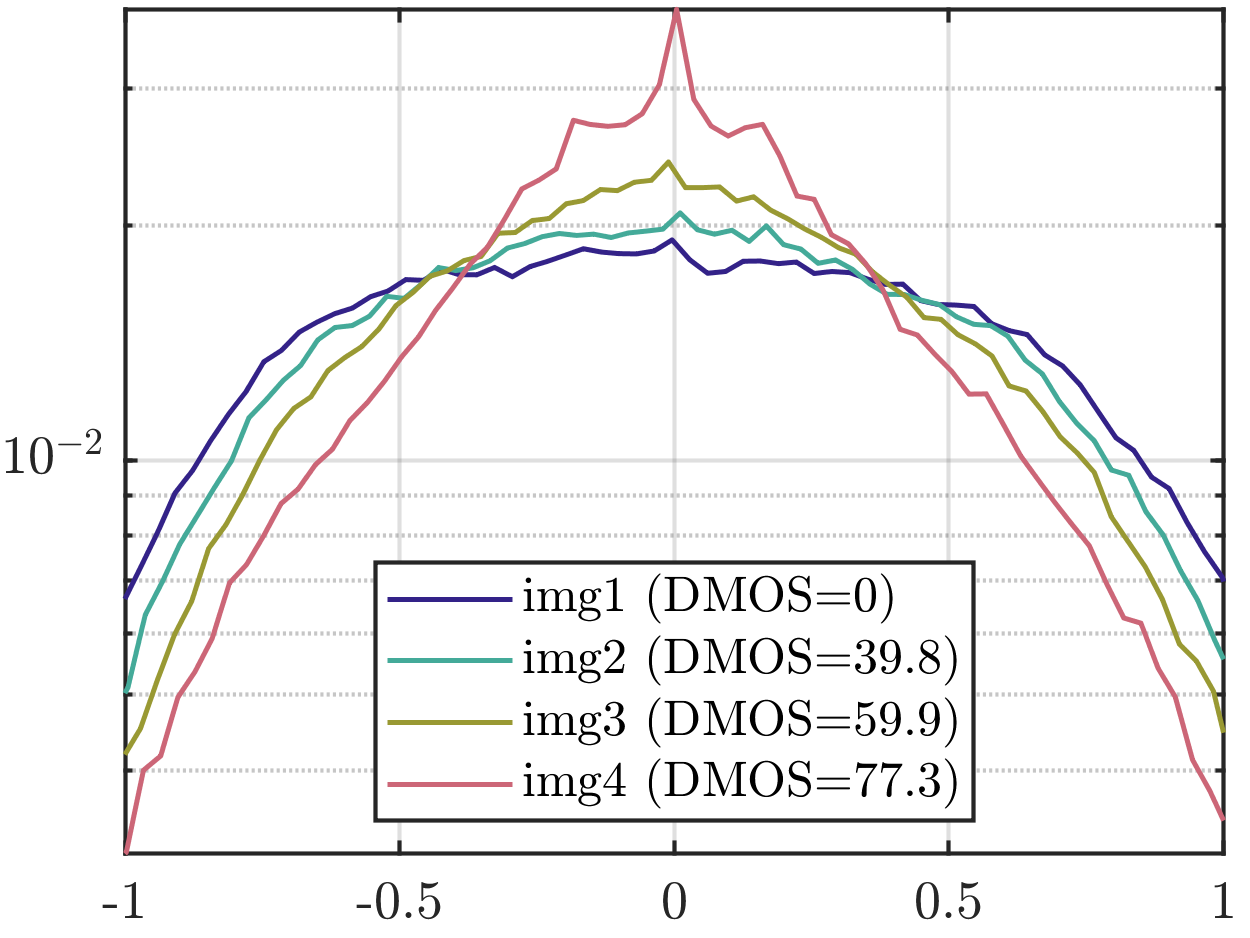} 
\end{tabular}
\caption{Histograms of MSCN coefficients of the four frames in Fig.~\ref{fig:sample-frame}.}
\label{fig:mscn}
\end{figure}

To illustrate how compression impairments affect NSS features derived from MSCN coefficients, we selected a group of videos from the LIVE-VQA\cite{5404314} dataset, specially the sequence \texttt{Park Run (PR)} having bitrates varied by H.264 compression from 200 kbps to 5 Mbps. Representative frames are shown in Fig.~\ref{fig:sample-frame}. Fig.~\ref{fig:mscn} plots the histograms of the MSCN coefficients of these frames as the compression distortion is varied. Showing clear differences in the profiles which may be expressed as parametric differences, suggesting that parameters estimated from the GGD distributions are reliable indicators of perceptual quality on videos distorted by H.264 compression.

\begin{figure}[!t]
\centering
\footnotesize
\def\xheight{0.485}
\setlength{\tabcolsep}{1pt}
\begin{tabular}{cc}
\includegraphics[width=\xheight\linewidth]{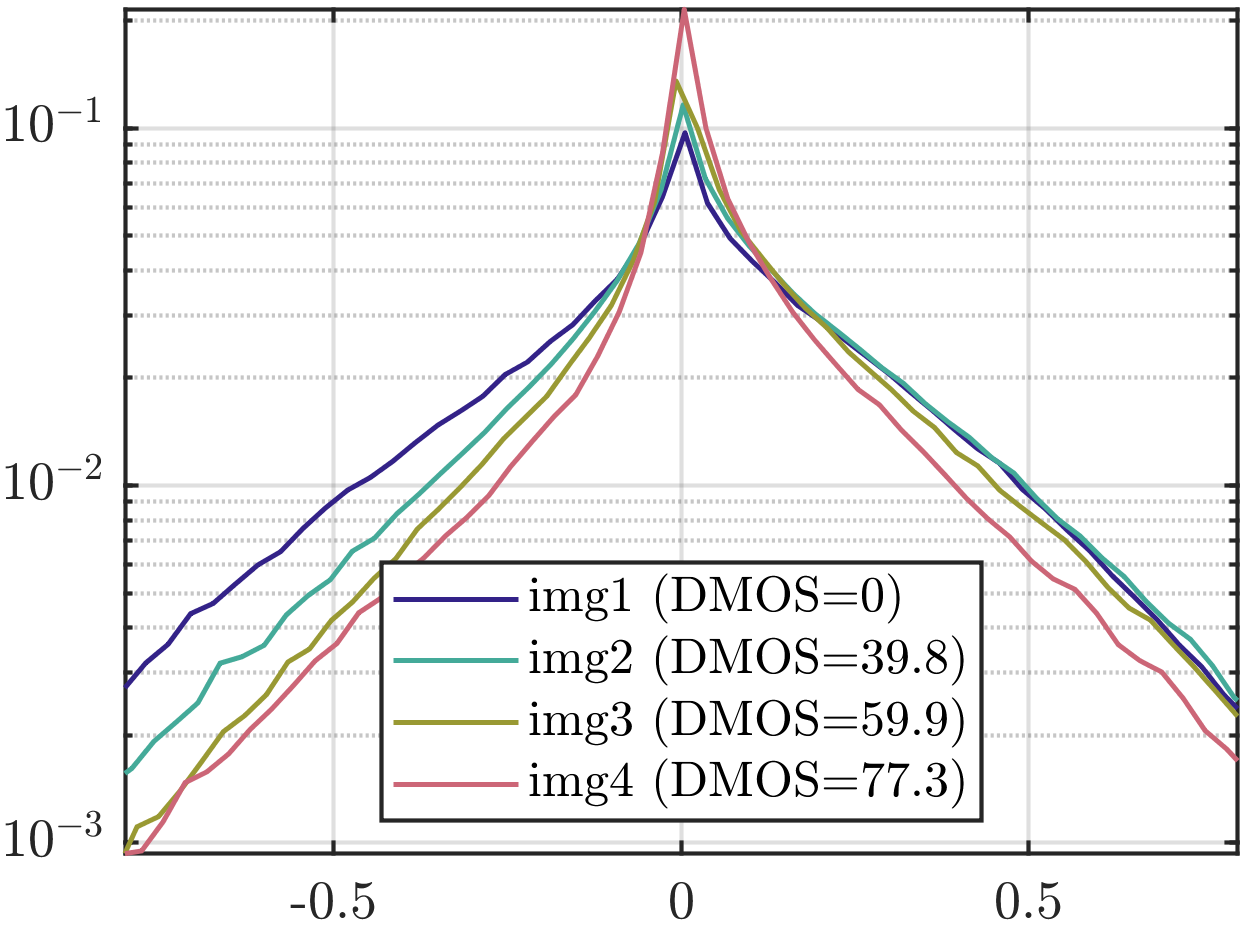} &
\includegraphics[width=\xheight\linewidth]{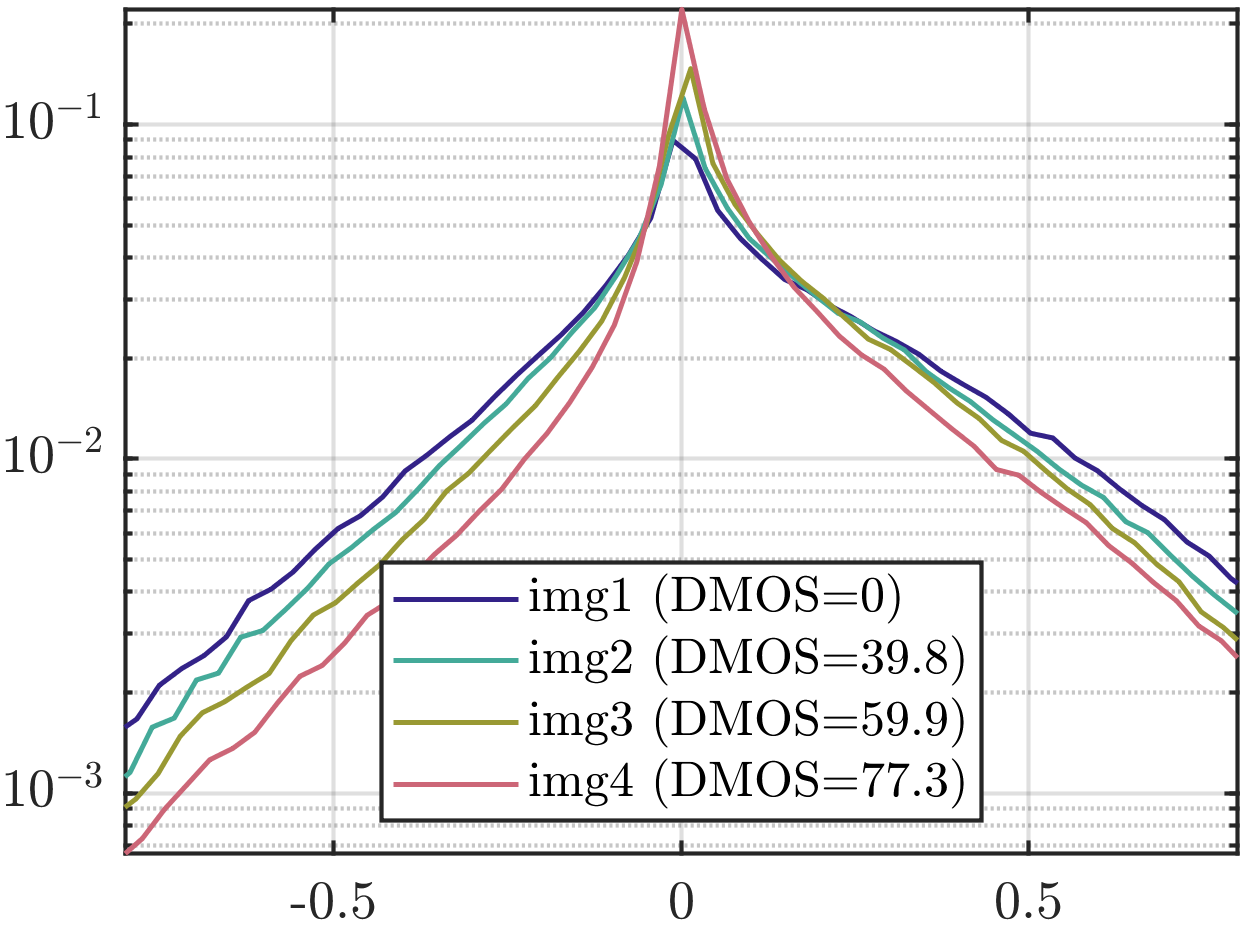} \\
\includegraphics[width=\xheight\linewidth]{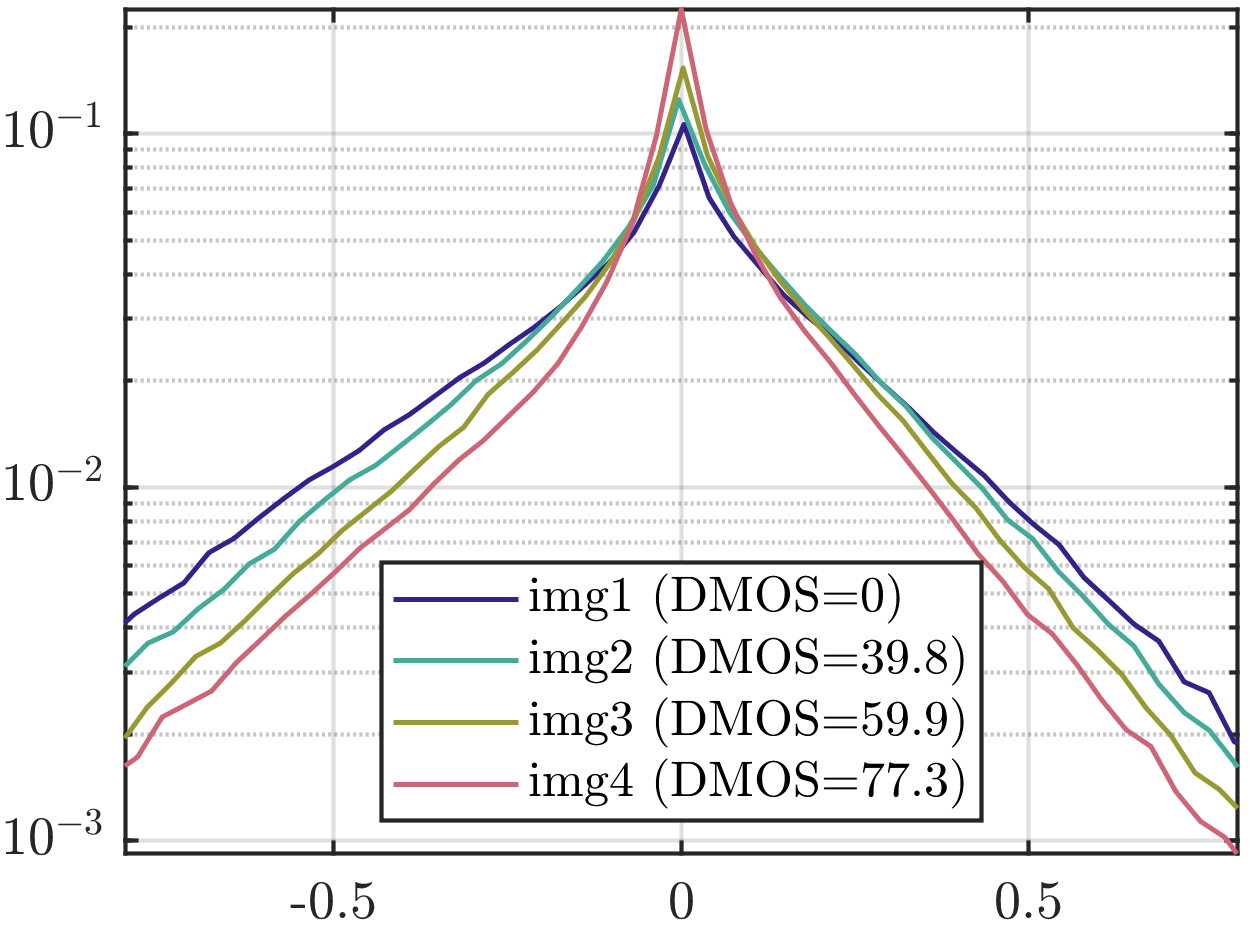} &
\includegraphics[width=\xheight\linewidth]{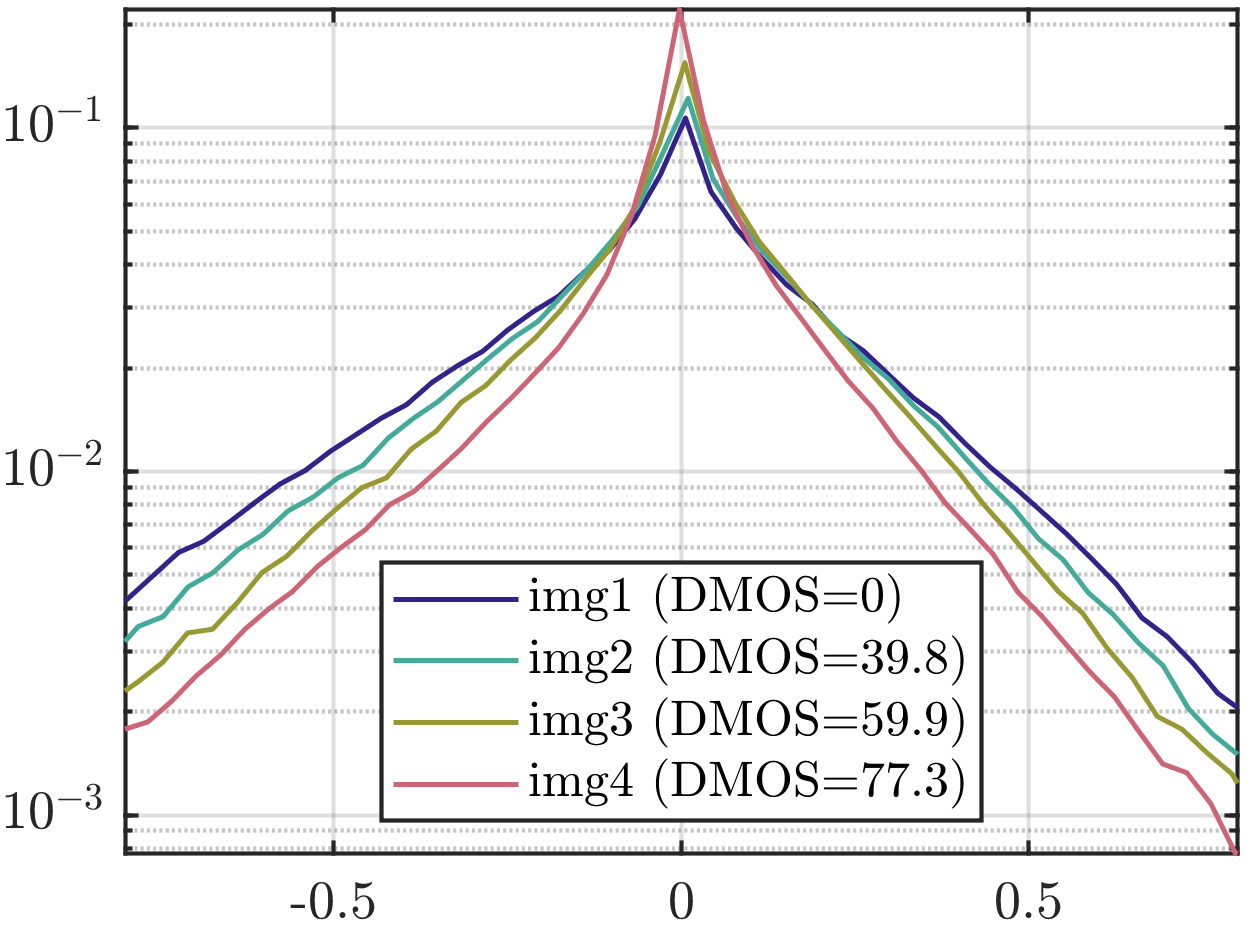}
\end{tabular}
\caption{Histograms of products of MSCN coefficients along four orientations $(H,V,D1,D2)$ on the four frames in Fig.~\ref{fig:sample-frame}.}
\label{fig:mscn-pair}
\end{figure}

Fig.~\ref{fig:mscn-pair} plots the histograms of products of adjacent pairs of the MSCN coefficients of videos compressed at different levels. These plots also illustrate characteristics variations in the product MSCN distributions. Fitting these curves with the AGGD yields parameters that capture modifications of the spatial correlations structure of videos introduced by compression.

\subsubsection{Gradient and Laplacian Features}
Using the unsmoothed gradient makes it possible to capture very fine details that may affect perceived quality. We estimate the gradient magnitude of each video frame using the Sobel kernel with horizontal and vertical operators
\begin{equation}
    H_x = \begin{bmatrix}
        1 & 0 & -1 \\
        2 & 0 & -2 \\
        1 & 0 & -1 \\
    \end{bmatrix},
\quad
    H_y = \begin{bmatrix}
         1 & 2 & 1 \\
         0 & 0 & 0 \\
        -1 & -2 & -1 \\
    \end{bmatrix}.
\end{equation}
The GM of each video frame is then calculated as
\begin{equation}
    GM = \sqrt{(I*h_x)^2+(I*h_y)^2},
\end{equation}
where $'*'$ denotes discrete convolution.
\par 
Coarser details are captured using the smoothed twice-derivative Laplacian-of-Gaussian (LOG) operator, which is calculated as:
\begin{equation}
    LoG = I * h_{LoG},
\end{equation}
where the LOG kernel is defined as:
\begin{equation}
\label{eq:log-deri}
\begin{gathered}
    h_{LOG}(x,y|\sigma) = \frac{\partial ^2}{\partial x^2}g(x,y|\sigma) + \frac{\partial ^2}{\partial y^2}g(x,y|\sigma) \\
    = \frac{1}{2\pi\sigma^2}\frac{x^2+y^2-2\sigma^2}{\sigma^4}exp(-\frac{x^2+y^2}{2\sigma^2}),
\end{gathered}
\end{equation}
where $g_\sigma(x,y)$ is an isotropic Gaussian function and $\sigma$ is a scale parameter. The window size is $9\times 9$.

\begin{figure*}[!t]
\centering
\footnotesize
\def\xheight{0.3}
\setlength{\tabcolsep}{2pt}
\begin{tabular}{ccc}
\includegraphics[width=\xheight\linewidth]{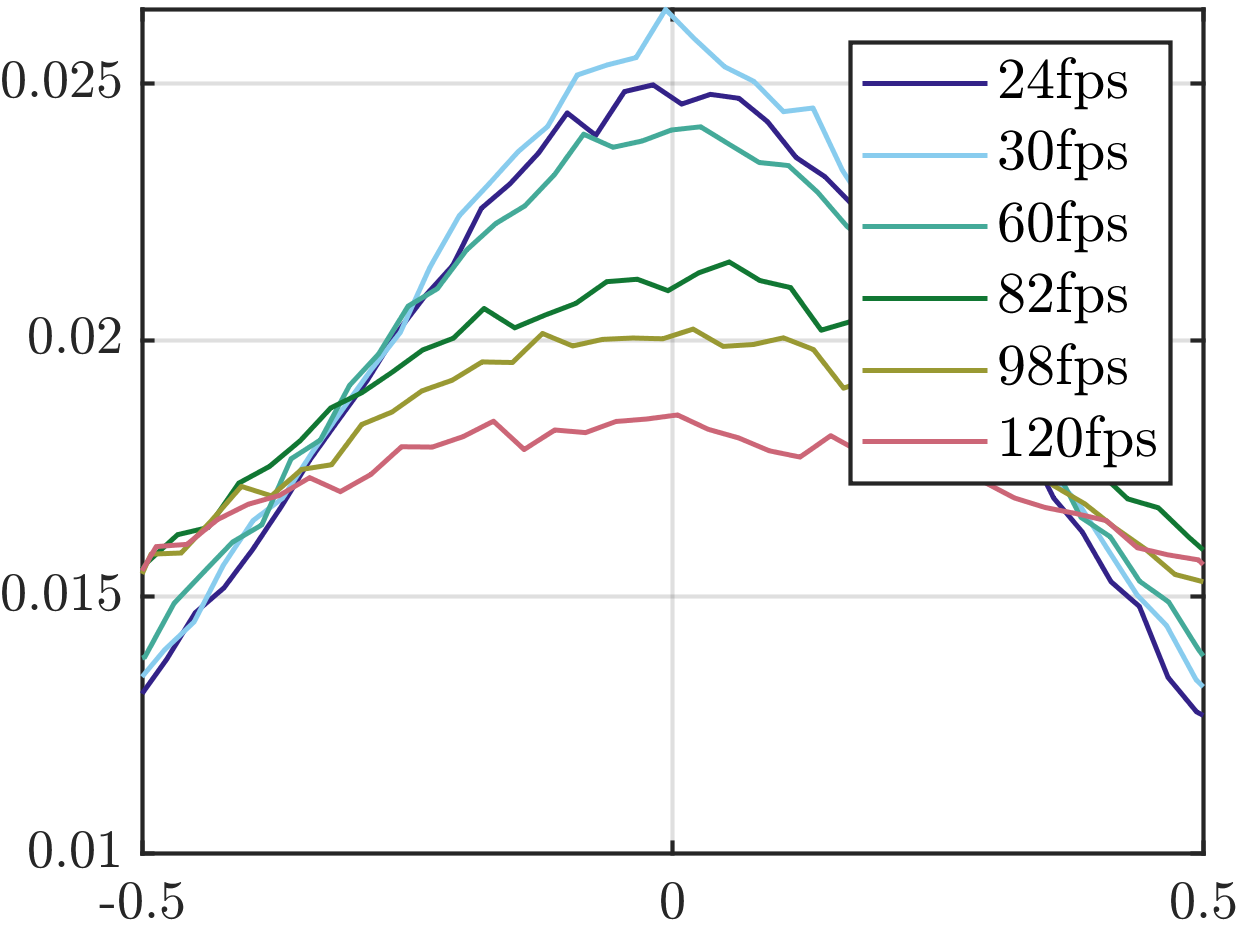} &
\includegraphics[width=\xheight\linewidth]{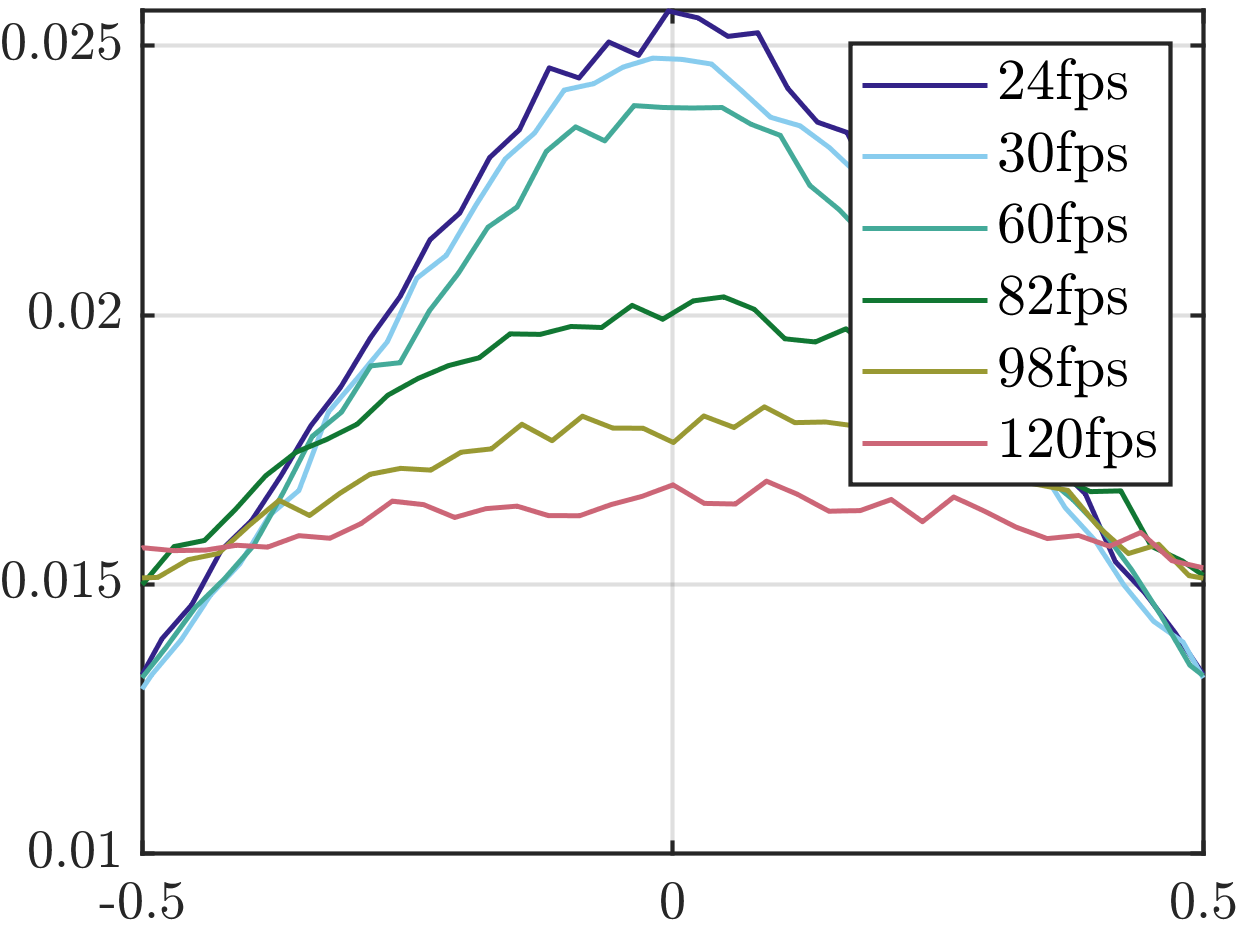} &
\includegraphics[width=\xheight\linewidth]{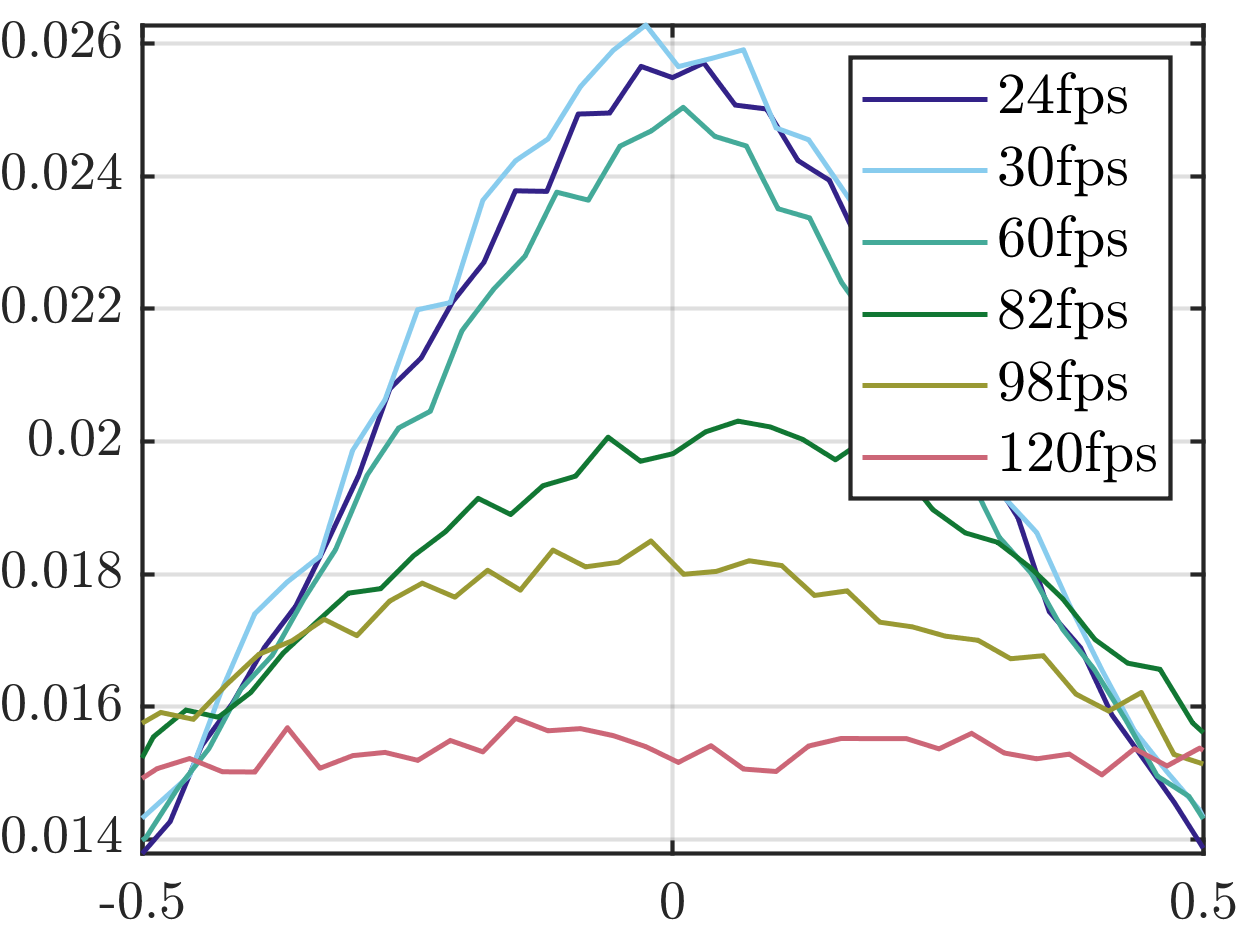} 
\\
(a) \texttt{Library} using Haar &
(b) \texttt{Library} using Db2 &
(c) \texttt{Library} using Bior22
\\\includegraphics[width=\xheight\linewidth]{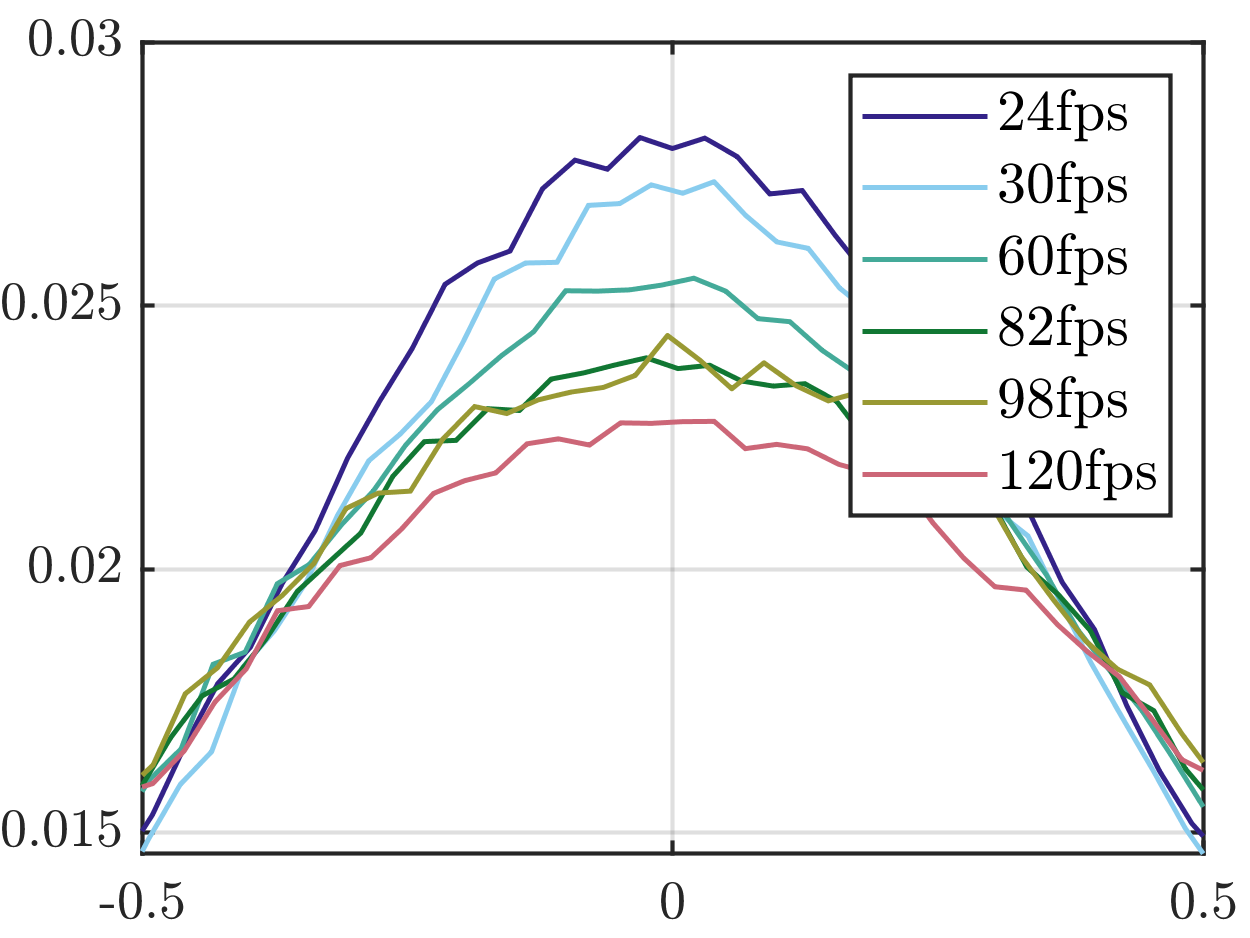} &
\includegraphics[width=\xheight\linewidth]{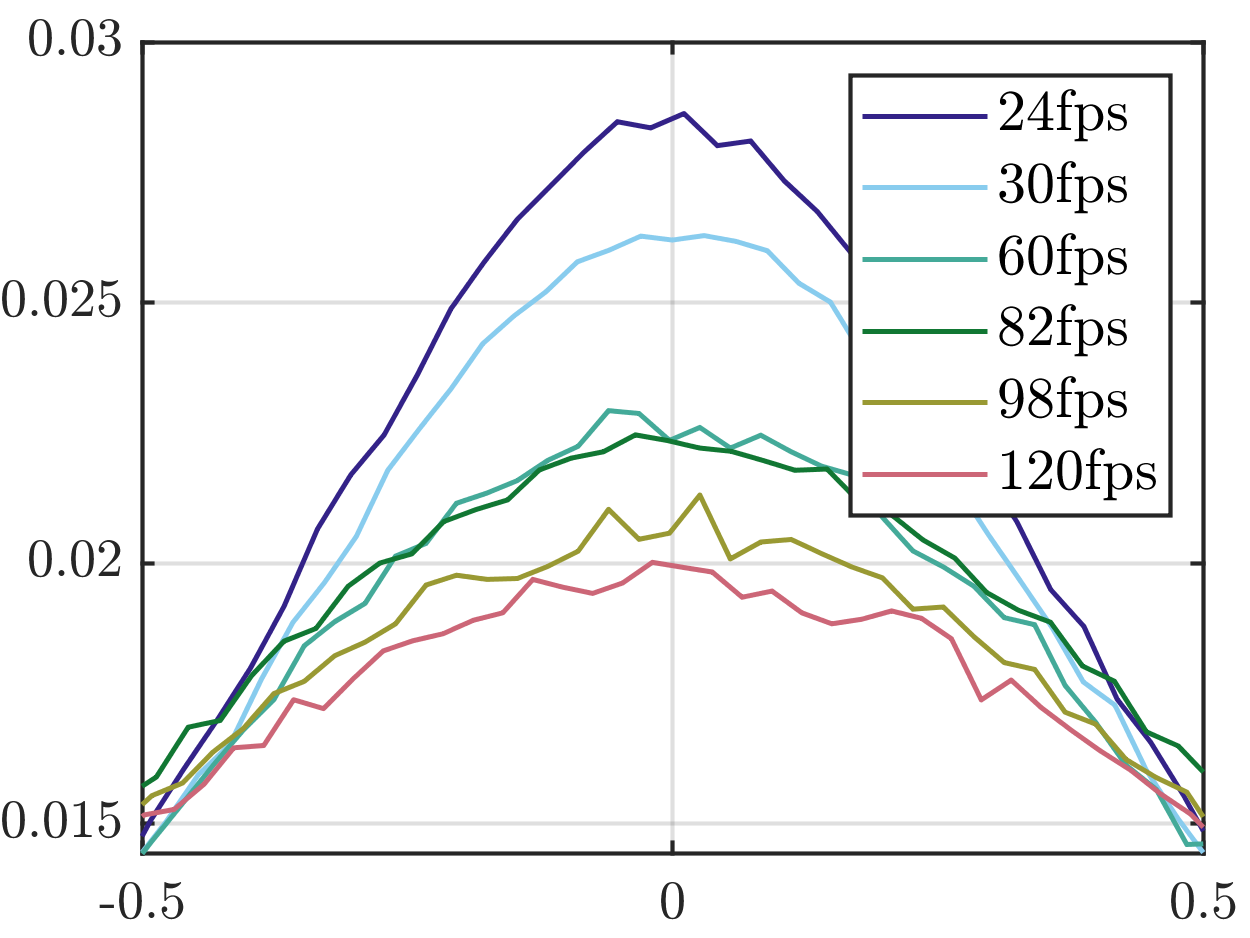} &
\includegraphics[width=\xheight\linewidth]{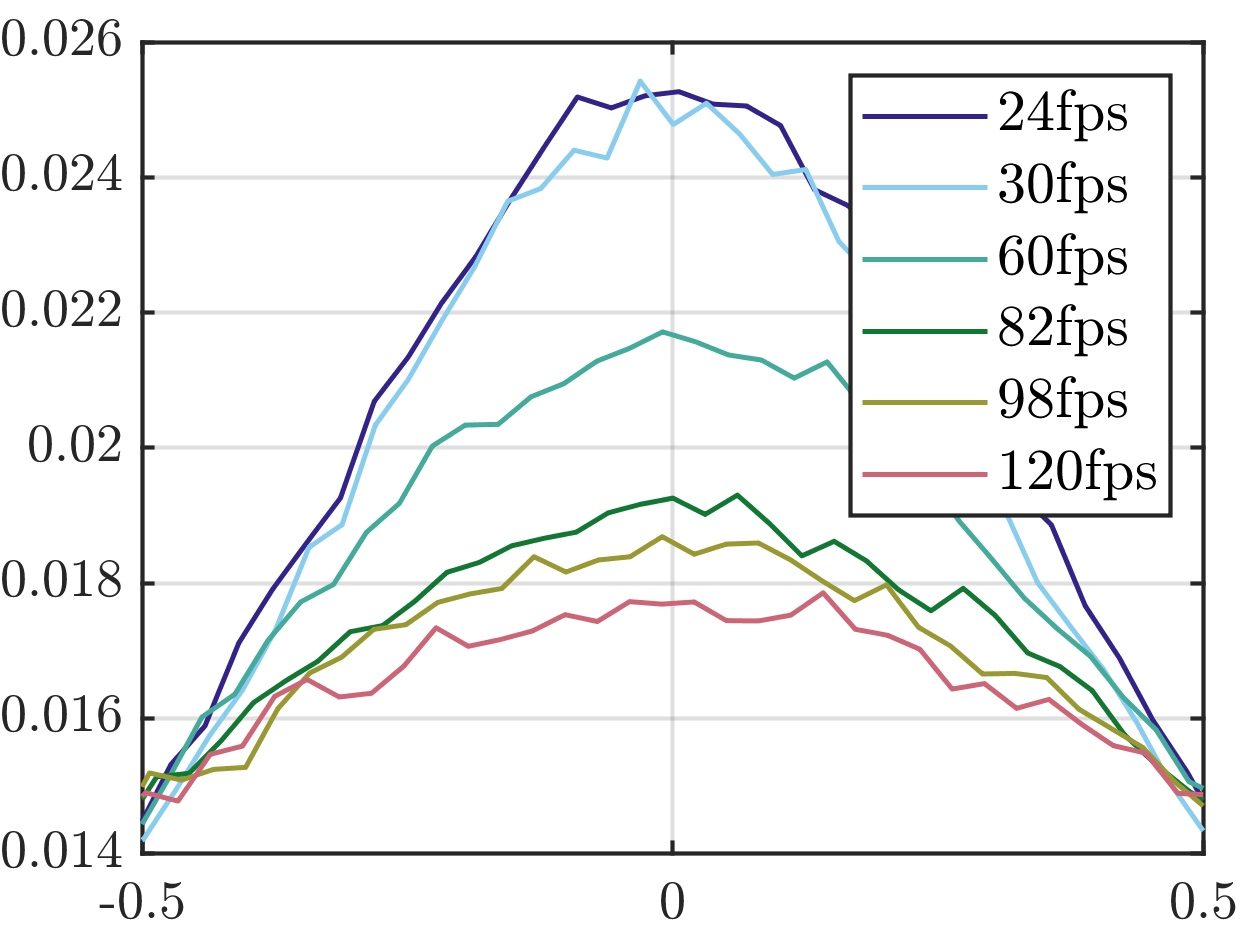} 
\\
(d) \texttt{Bouncyball} using Haar &
(e) \texttt{Bouncyball} using Db2 &
(f) \texttt{Bouncyball} using Bior22
\end{tabular}
\caption{Empirical distributions (histograms) of MSCN normalized coefficients of the \textbf{fourth} subband of the temporal filtered responses on sequences \texttt{Library} ((a)-(c)) and \texttt{Bouncyball} ((d)-(f)) from the LIVE-YT-HFR database. Each plot shows histograms of the same bandpass-filtered contents but at different framerates: 24, 30, 60, 82, 98, 120 fps. The top row shows the distributions of bandpass coefficients using (a) \textbf{Haar}, (b) \textbf{Db2}, and (c) \textbf{Bior22} filters on the sequence \texttt{Library}, while the bottom row shows the distrbutions using the same types of filters on the sequence \texttt{Bouncyball}.}
\label{fig:temp_stats}
\end{figure*}

The overall flow of spatial feature extraction is as follows. The YUV components of each frame are fed into the basic NSS feature extractor to obtain low-frequency statistical measurements. The GM and LOG are applied on the Y channel to capture the high-frequency and mid-frequency information descriptive of each frame. The five image maps $(Y, U, V, GM, LOG)$ are then each processed by the NSS-34 feature extraction module, yielding a total of $170$ spatial quality-aware features.

Since videos are naturally multiscale, so are distortions and perception of them. Therefore, as in prior NSS-based models~\cite{6353522,FRIQUEE} we compute NSS features on the $(Y, U, V)$ components of two scales (the original- and half-scale) while the features are computed on $(GM, LOG)$ only at the half scale. The half-scale is obtained by down-sampling by 2 after smoothing by a small Gaussian kernel. Overall, $272$ spatial features are obtained.

\subsection{Temporal Feature Design}
\label{ssec:temporal-feature}

Since our goal is to design models of blind video quality as affected by frame rate variations, we are interested in no-reference perceptual quality-aware features that capture relevant temporal aspects of VFR-induced distortions. However, existing BVQA models have only deployed simple statistics of adjacent frame-difference~\cite{sinno2019spatio, mittal2015completely, saad2014blind}, or of expensive between-frame motion vectors~\cite{korhonen2019two, seshadrinathan2009motion}. As we shall see, these prior models perform poorly on VFR distortions, which generally develop over multiple frames. Therefore, inspired by the effective use of temporal bandpass filters in the FR model ST-GREED~\cite{9524440}, we take a similar approach, but without access to a reference signal, to build NR-VQA models sensitive to VFR distortions.

Consider a bank of $K$ 1D purely temporal bandpass filters denoted as $b_k, k \in \{0,...,K-1\}$, where $k$ is the subband index. The temporal bandpass responses of these filters to a video $F(\mathbf{x},t)$ (where $\mathbf{x} = (x,y)$ and $t$ are spatial and temporal coordinates, respectively) is 
\begin{equation}
\label{eq:bandpass filter}
    B_k(\mathbf{x},t) = V(\mathbf{x},t)*b_k(t), \quad k \in {1,...,K},
\end{equation}
where $*$ is the discrete convolution operator, and $B_k$ is the response of $b_k$. Of course, frame-differencing is a special case of (\ref{eq:bandpass filter}). We compared three types of wavelets in this paper: Haar, Daubechies-2 (DB2), and Biorthogonal-2.2 (bior22). These are the same temporal filters used to create different versions of ST-GREED~\cite{9524440}.

To illustrate the temporal bandpass statistics of videos having different frame rates, we used the sequences \texttt{Library} and \texttt{Bouncyball} from LIVE-YT-HFR with six variants of framerate (24, 30, 60, 82, 98, 120 fps). Fig.~\ref{fig:temp_stats} plots the empirical distributions of the spatial MSCN-normalized responses of these videos after temporal Haar, Db2 and Bior22 filtering. As may be seen, the framerate affects the shape and spread of each video's histograms in a systematic way. This suggests that all three choices of wavelet filter are able to capture statistical differences between video frame rates that may yield quality-aware features that can account for the perception of VFR distortions. 

Following the above observations, we developed three algorithms in parallel, depending on the bandpass filters used. Haar filters are simple and compute efficient, and often perform surprisingly well. Db2 filters possess excellent representation power, as do the Bior22 with the additional advantage of symmetry. For each filter type, we model the spatial MSCN distributions of the aggregate space-time subband responses using GGD and AGGD (product) models, thereby extracting quality-aware statistical features on all MSCN coefficients for all seven temporal subbands. As in the preceding, this feature extraction process is applied at two spatial scales (original and half scale), yielding a total of 476 temporal features (34 features/band)×(7 subbands)×(2 scales)=476) at each time step. All of the computed features are then averaged across all frames.

\subsection{Summary and Learning of Features}
\label{ssec:integration}

We summarize the overall flow of FAVER in the following. Given an input video $F(\mathbf{x},t)$, uniformly sample one frame per second, then apply the basic spatial feature extraction module (Sec.~\ref{ssec:Spatial-feature}) to the $(Y,U,V)$ components at two scales, as well as on the feature maps $(GM, LOG)$ at the reduced-scale only. All the extracted spatial features are then average pooled across frames, yielding a 272-dimensional spatial feature vectors. Then the basic 34-dim feature extractor module is applied to the temporal subband responses using each of Haar/Db2/Bior22 filter sets (as three different realizations) on which spatial MSCN is applied, to extract temporal features at two scales. Since purely temporal features capture statistical behavior over time, which is less sensitive to fine spatial details, the videos were spatially downscaled to have heights of 512 pixels (maintaining the aspect ratio) to accelerate computation. We have observed that this does not affect prediction performance. Combining the 272 spatial features and the 476 temporal features yields a total of 748 FAVER features.

Most feature-based BVQA models train a single support vector machine (SVM) to perform feature-to-score mapping. However, we observed that the spatial and temporal features in FAVER have different distributions, hence we adopted ensemble learning~\cite{bampis2018spatiotemporal, 1688199} to better exploit the bifurcated space/time statistical structure. Specifically, we trained two separate SVM models, one to learn spatial features, and the other to learn temporal features. FAVER then simply averages the spatial and temporal predictions to produce a video quality final prediction. We observed that this ensemble scheme can often reliably improve performance on relatively small datasets like~\cite{madhusudana2021subjective,8531714}.

\section{Experimental Results}
\label{sec:experimental}

\subsection{Evaluation Protocol}
\label{ssec:evaluation-protocol}

\subsubsection{Dataset Benchmarks}
\label{sssec:dataset-benchmarks}

We conducted comparisons of the performances of FAVER and other IQA/VQA models on the only available HFR video quality dataset, LIVE-YT-HFR~\cite{madhusudana2021subjective}, that contains both VFR and compression distortions. We also conducted experiments using BVI-HFR~\cite{8531714}, which contains VFR content but without compression variations. The main characteristics of LIVE-YT-HFR and BVI-HFR are summarized in Table~\ref{table:datasets}.

\begin{table}[!t]
\setlength{\tabcolsep}{3pt}
\centering
\footnotesize
\caption{Metadata of the VQA databases used in the experiments.}
\label{table:datasets}
    \begin{tabular}{lcccccc}
    \toprule
    \textbf{Database} & \begin{tabular}[c]{@{}c@{}}\textbf{\#Source}\\ \textbf{videos}\end{tabular} & \begin{tabular}[c]{@{}c@{}}\textbf{Framerate} \\ \textbf{(fps)}\end{tabular}   & \textbf{CRF}                   &
    \textbf{Distortion} &
    \begin{tabular}[c]{@{}c@{}}\textbf{\#Total}\\ \textbf{videos}\end{tabular} \\ \midrule
    BVI-HFR~\cite{8531714}  & 22                                                             & \begin{tabular}[c]{@{}c@{}}15, 30, \\60, 120\end{tabular}                                                    & None    & Framerate & 88                                                     \\  
    LIVE-YT-HFR~\cite{madhusudana2021subjective} & 16                                                             & \begin{tabular}[c]{@{}c@{}}24, 30, 60, \\ 82, 98, 120\end{tabular} & 0-63                           &                                                 \begin{tabular}[c]{@{}c@{}}Framerate, \\ Compression\end{tabular}         & 480                                                    \\ 
    \bottomrule
    \end{tabular}
\end{table}

\subsubsection{BVQA Model Baselines}
\label{sssec:bvqa-models}

We studied the performance of FAVER against those of several popular and widely used BVQA models: BRISQUE~\cite{mittal2012no}, GM-LOG~\cite{xue2014blind}, HIGRADE~\cite{kundu2017no}, FRIQUEE~\cite{FRIQUEE}, CORNIA~\cite{6247789}, HOSA~\cite{7501619} and NIQE~\cite{6353522}. As in ~\cite{9405420, DBLP}, features or scores computed using each of these IQA models were computed at a rate of one frame per second, then averaged across all frames to obtain final video features or scores. We also studied several leading BVQA models: V-BLIINDS~\cite{6705673}, VIDEVAL~\cite{9405420}, RAPIQUE~\cite{DBLP} and TLVQM~\cite{korhonen2019two}.

\begin{table}[!t]
\centering
\footnotesize
\caption{Performance comparison of evaluated models on the LIVE-YT-HFR database. The boldfaced entries indicate the top three performers for each performance metric.}
\label{table:performance-live-yt-hfr}
\begin{tabular}{l|ccc}
\toprule
\textbf{Model}                    &  \textbf{SROCC↑}                      &  \textbf{PLCC↑}                       & \textbf{RMSE↓}                       \\ 
\midrule
NIQE\cite{6353522}                                     & 0.1371                                           & 0.4184                      & 11.0853                     \\ 
BRISQUE\cite{mittal2012no}                               & 0.3190                                           & 0.4196                      & 11.0615                     \\ 
GM-LOG\cite{xue2014blind}                                & 0.4950                                           & 0.6049                      & 9.6585                      \\ 
HIGRADE\cite{kundu2017no}                           & 0.4640                                         & 0.5557                      & 10.2196                     \\ 
FRIQUEE\cite{FRIQUEE}                              & 0.4801                                          & 0.5723                      & 10.0235                     \\ 
CORNIA\cite{6247789}                              & 0.3271                                        & 0.4954                      & 10.6968                     \\ 
HOSA\cite{7501619}                              & 0.4623                                           & 0.6043                      & 9.7896                      \\ 
VBLIINDS\cite{6705673}                             & 0.3917                                          & 0.4675                      & 10.8168                     \\ 
VIDEVAL\cite{9405420}                            & 0.4748                                           & 0.5665                      & 10.0547                     \\ 
TLVQM\cite{korhonen2019two}                                & 0.4295                                          & 0.5047                      & 10.5108                     \\ 
RAPIQUE\cite{DBLP}                               & 0.4566                                   & 0.5665                      & 9.8969                      \\ 
\midrule
FAVER-Haar             & \textbf{0.5864}                                         & \textbf{0.6546}                      & \textbf{9.1789}                      \\ 
FAVER-Db2                & \textbf{0.6195}                      & \textbf{0.6769}             & \textbf{8.9525}             \\ 
FAVER-Bior22              & \textbf{0.6350}                                          & \textbf{0.6923}                      & \textbf{8.7964}                      \\                  
\bottomrule
\end{tabular}
\end{table}

\begin{figure*}[!t]
\centering
\def\xheight{0.24}
\footnotesize
\subfigure[NIQE]{\includegraphics[width=\xheight\linewidth]{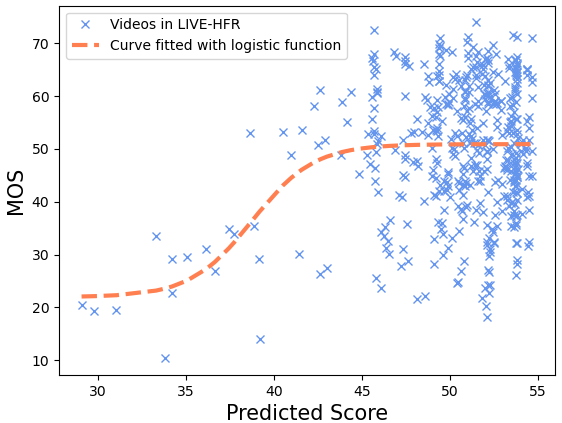}}
\subfigure[BRISQUE]{\includegraphics[width=\xheight\linewidth]{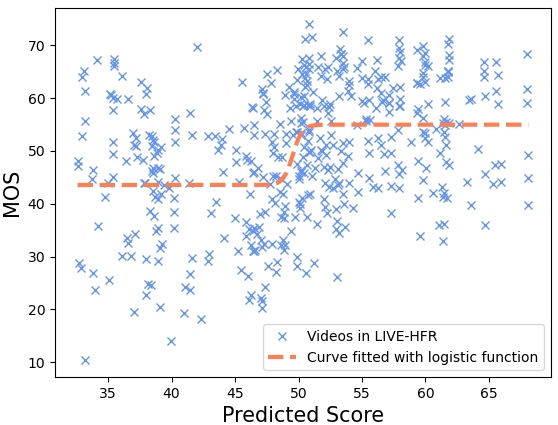}}
\subfigure[GM-LOG]{\includegraphics[width=\xheight\linewidth]{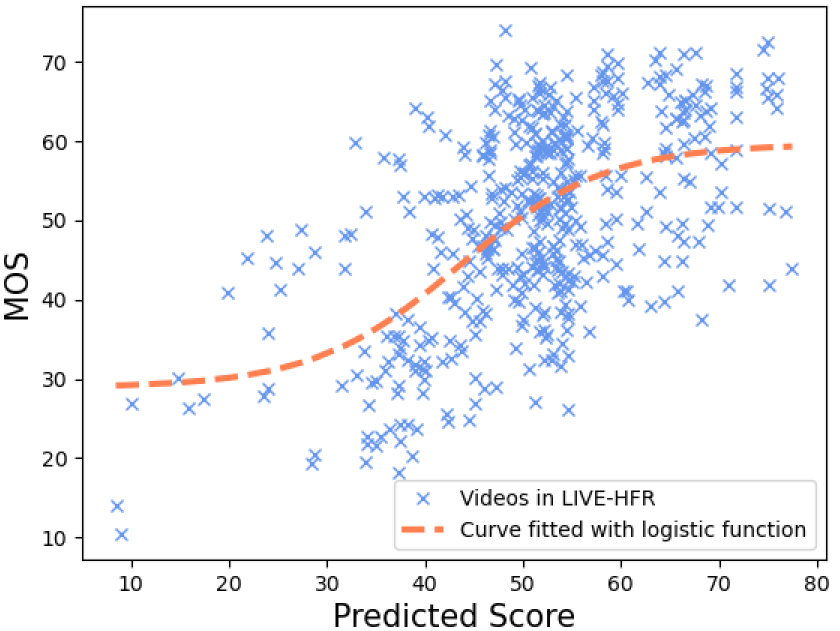}}
\subfigure[VBLIINDS]{\includegraphics[width=\xheight\linewidth]{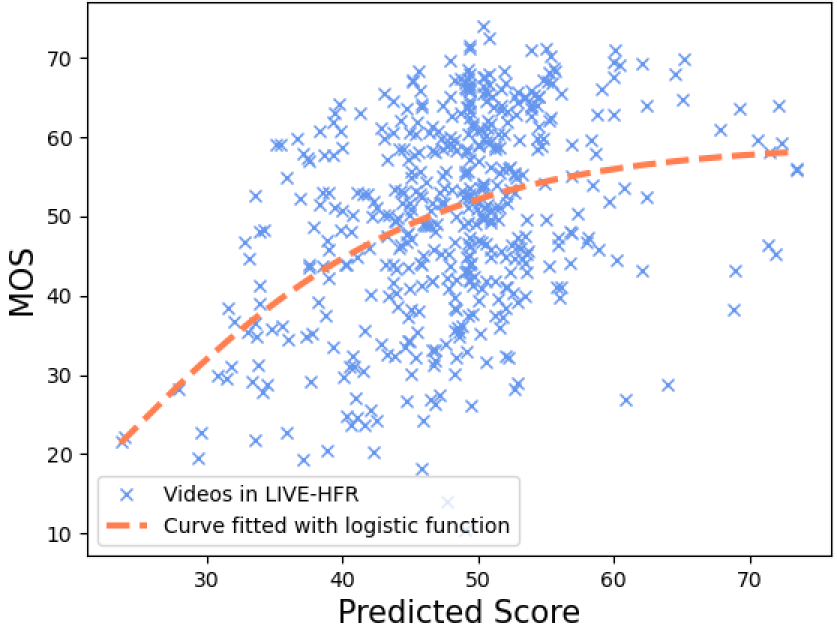}}
\\
\subfigure[VIDEVAL]{\includegraphics[width=\xheight\linewidth]{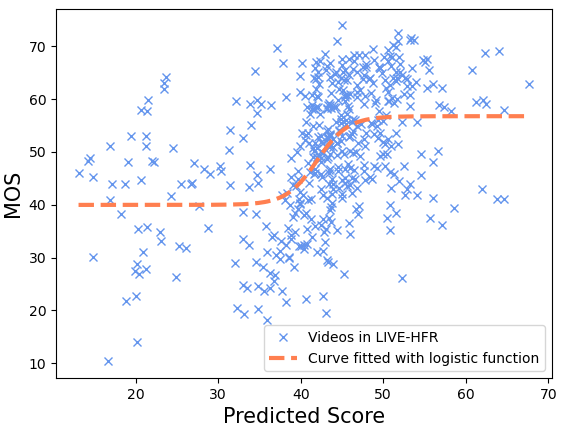}}
\subfigure[TLVQM]{\includegraphics[width=\xheight\linewidth]{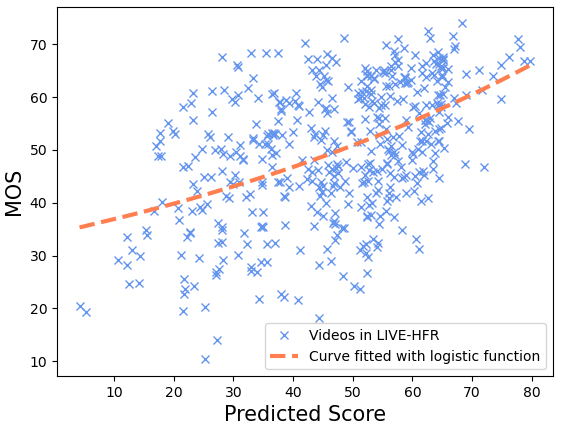}}
\subfigure[FRIQUEE]{\includegraphics[width=\xheight\linewidth]{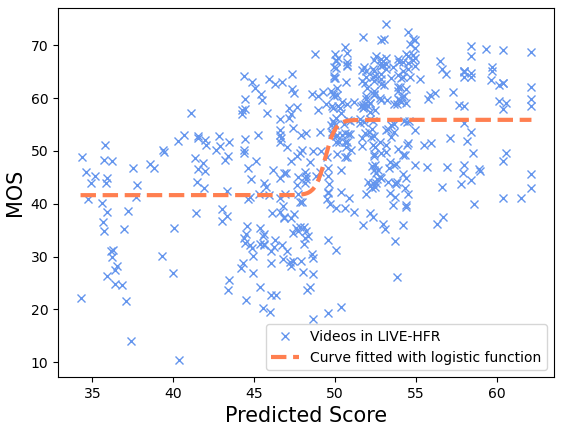}}
\subfigure[HIGRADE]{\includegraphics[width=\xheight\linewidth]{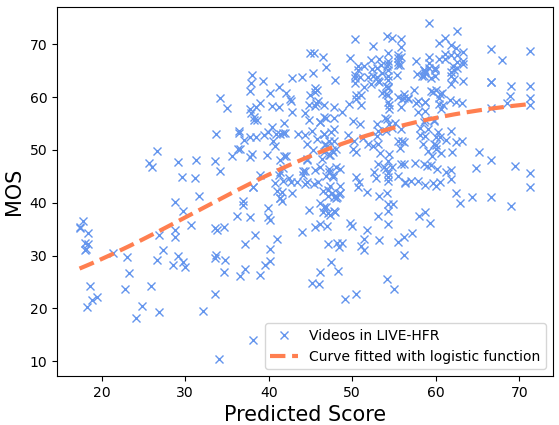}}
\\
\subfigure[RAPIQUE]{\includegraphics[width=\xheight\linewidth]{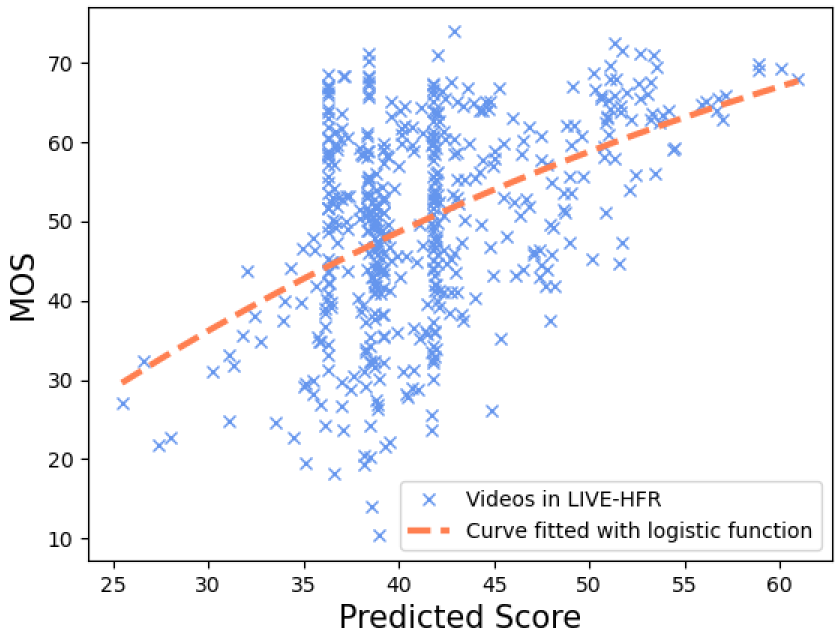}}
\subfigure[FAVER-Haar]{\includegraphics[width=\xheight\linewidth]{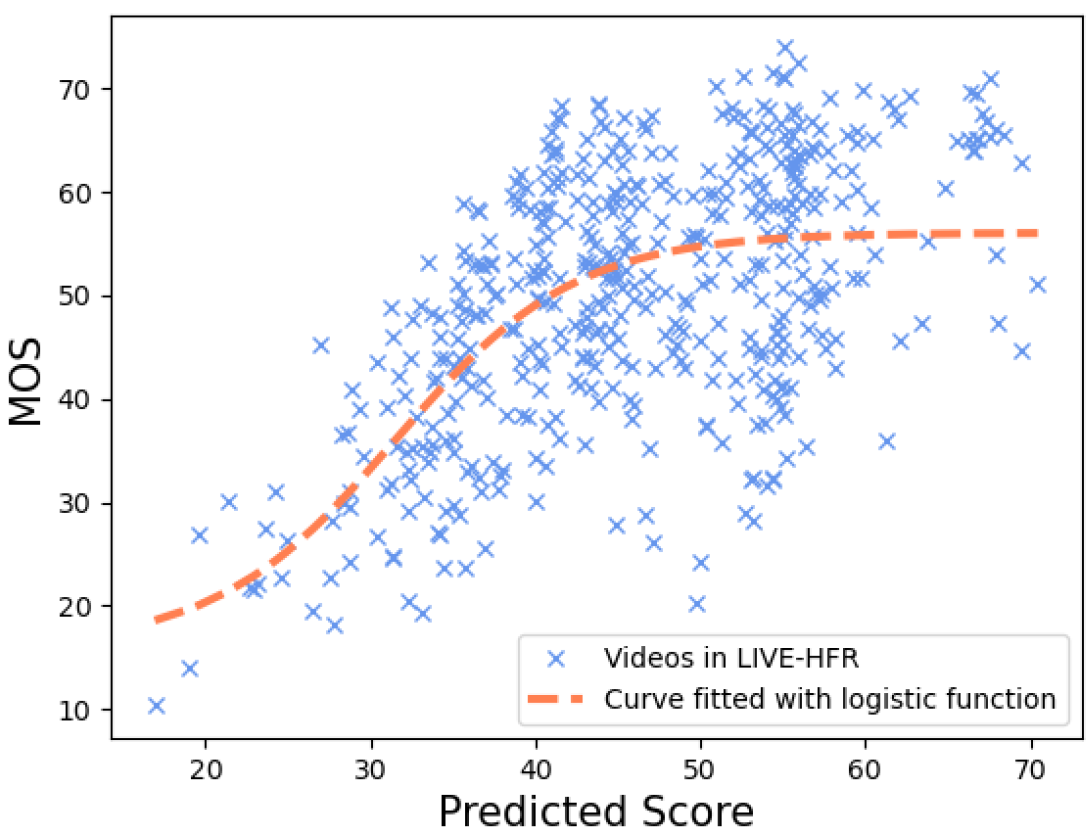}}
\subfigure[FAVER-Db2]{\includegraphics[width=\xheight\linewidth]{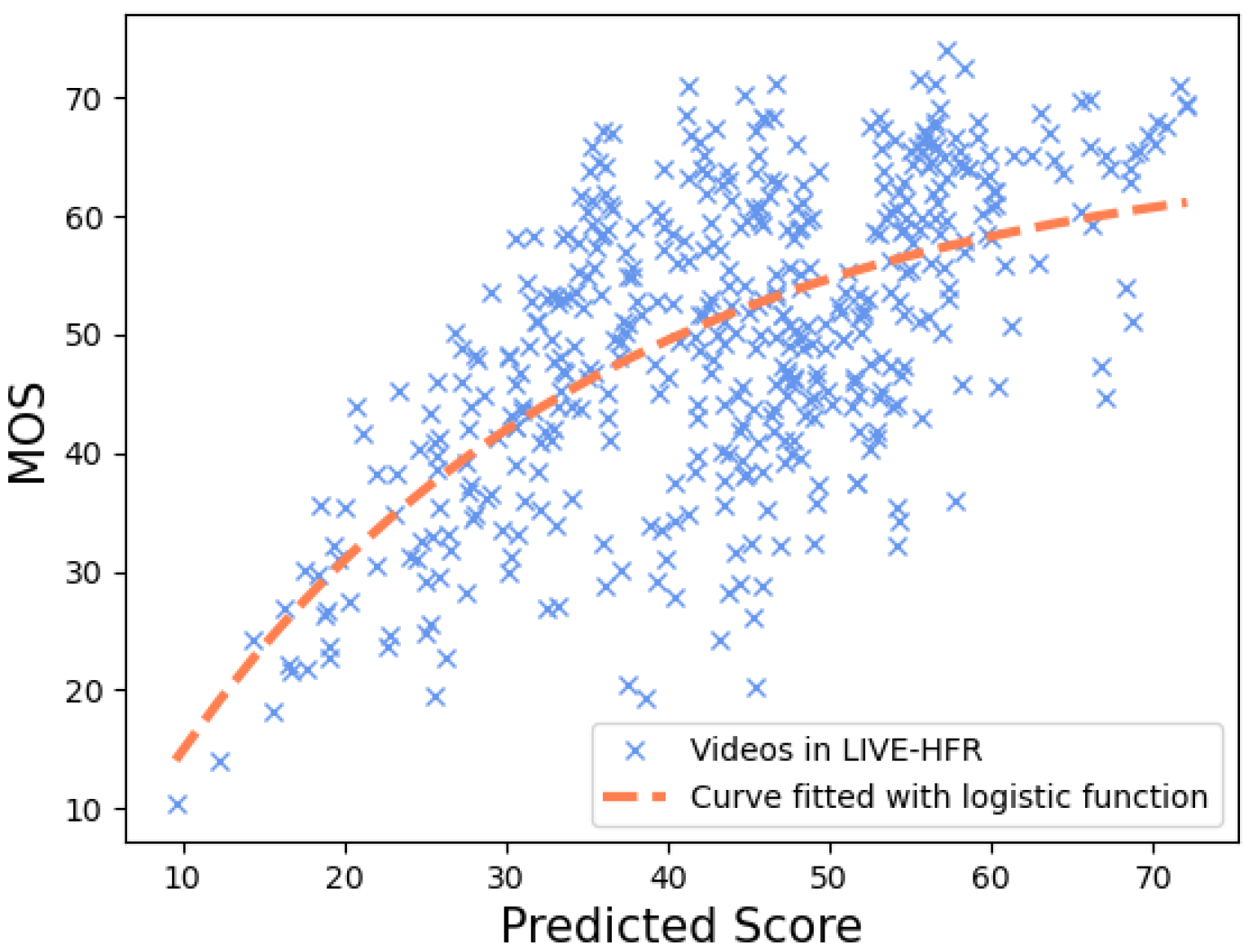}}
\subfigure[FAVER-Bior22]{\includegraphics[width=\xheight\linewidth]{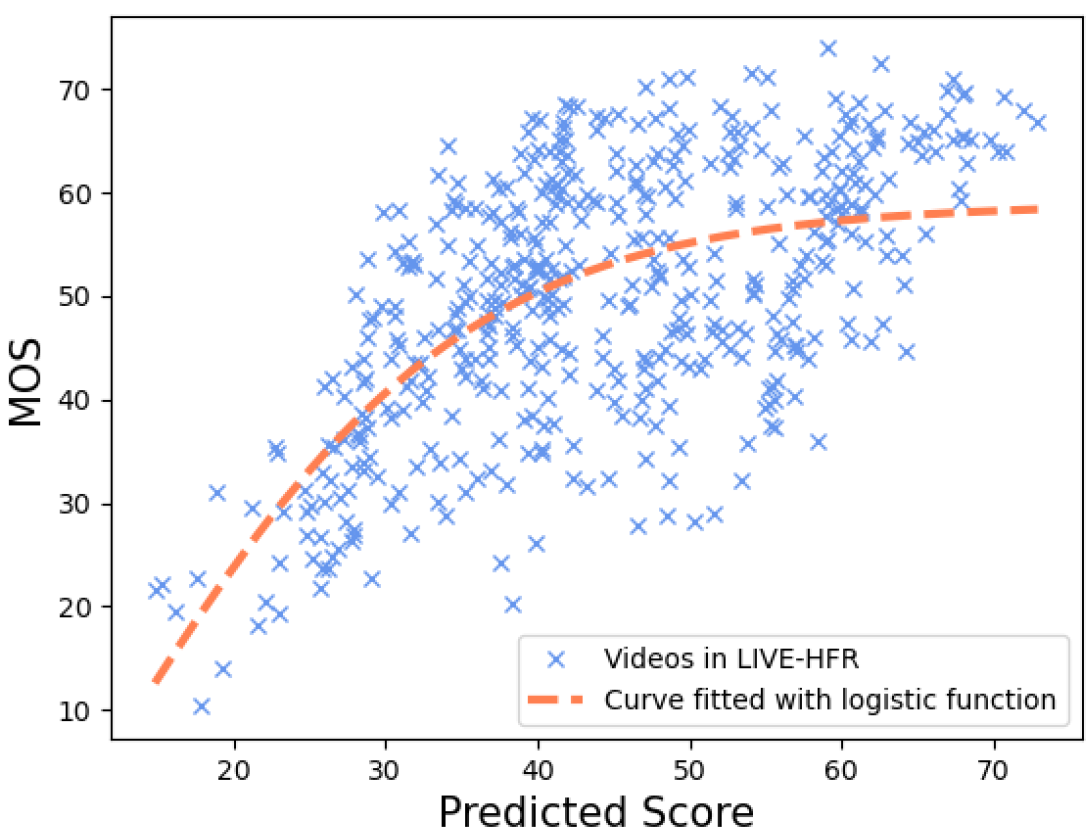}}
\caption{Scatter plots and logistic fitted curves of predictions versus MOS trained with a grid-search SVR using k-fold cross-validation on the LIVE-YT-HFR database for the following evaluated models: (a) NIQE, (b) BRISQUE, (c) GM-LOG, (d) V-BLIINDS, (e) VIDEVAL, (f) TLVQM, (g) FRIQUEE, (h) HIGRADE, (i) RAPIQUE, and FAVER family (j) FAVER-Haar, (k) FAVER-Db2, (l) FAVER-Bior22.}
\label{fig:scatter}
\end{figure*}

\begin{figure}[!t]
\centering
\footnotesize
\includegraphics[width=0.98\linewidth]{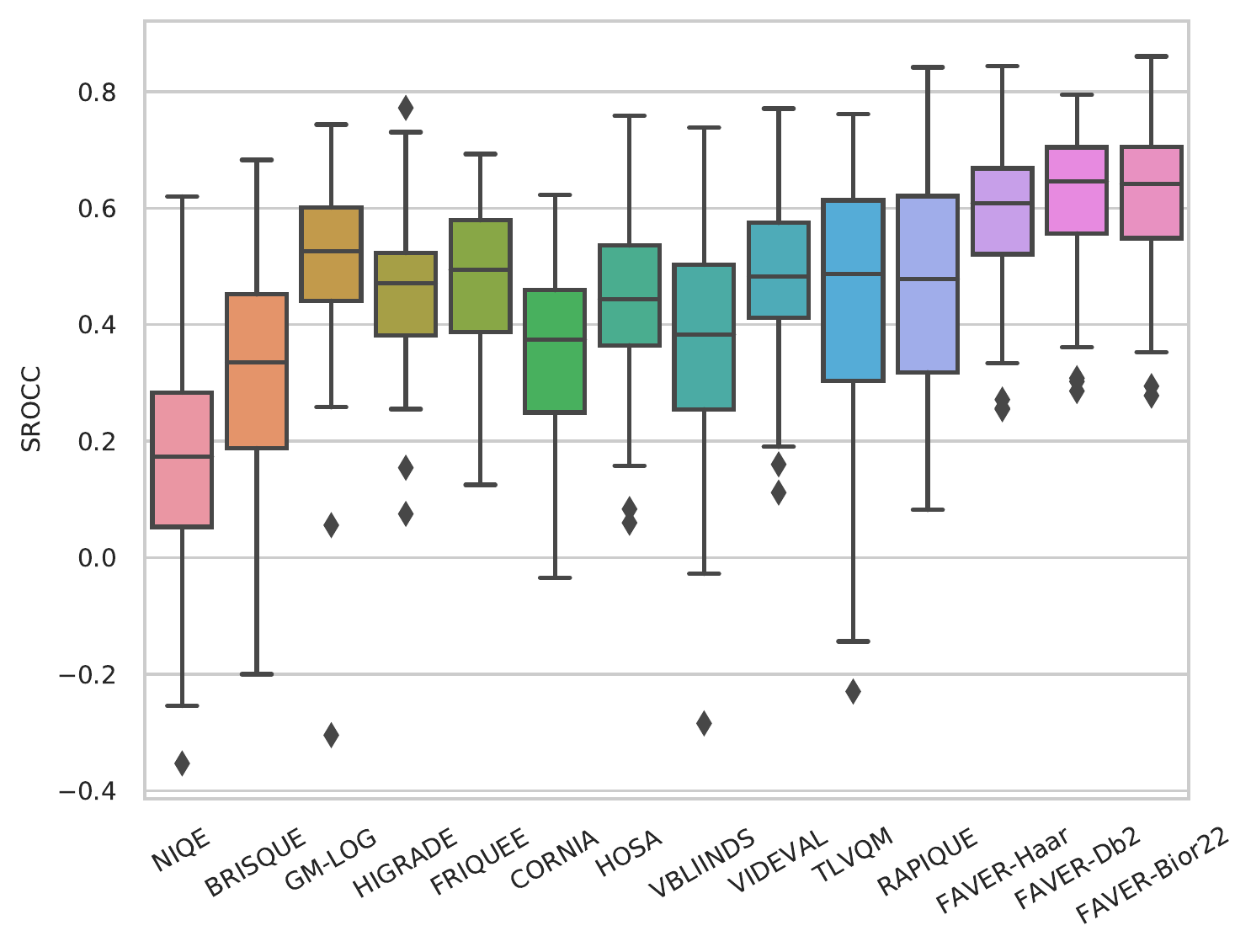} 
\caption{Boxplots of SROCC distributions over 100 iterations of train-test splits of multiple BVQA models evaluated on the LIVE-YT-HFR dataset.}
\label{fig:boxplots}
\end{figure}

\subsubsection{Regression and Evaluation Methods}
\label{sssec:Regression}

To optimize FAVER, we used the support vector regressor (SVR) with radial basis function (RBF) kernel as the back-end regression model to learn mappings from features to human opinion scores. A linear SVR is utilized in models containing more than one thousand features, to avoid overfitting to some extent. The ensemble system introduced in~\ref{ssec:integration} is trained throughout the learning process. The SVR parameters $(C,\gamma)$ and LinearSVR parameters $(C,\epsilon)$ were optimized via a randomized search on the training set. Following accepted protocols~\cite{6705673,9405420,korhonen2019two,9524440,DBLP,mittal2012no,6353522,FRIQUEE,xue2014blind,mittal2015completely}, we randomly subdivided the database into training and test sets comprising approximatedly 80\% and 20\% of the data, respectively. We repeated the train-test process over 100 random divisions, over which we report median values of all the performance metrics. On the LIVE-YT-HFR and BVI-HFR datasets, the exact ratios of training/test data on each iteration were 13:3 and 17:5, respectively, while guaranteeing that the training and test sets shared no versions (distorted or otherwise) of any original content. The performance metrics we employed are the Spearman’s rank-order correlation coefficient (SROCC), Pearson’s linear correlation coefficient (PLCC), and the root mean squared error (RMSE). SROCC evaluates the monotonicity of prediction performance, while PLCC and RMSE measure the prediction accuracy. Note that PLCC and RMSE were computed after performing a nonlinear four-parametric logistic regression to linearize the objective predictions to be on the same scale as MOS \cite{seshadrinathan2010study}:
\begin{equation}
\label{eq:logistic}
f(x)=\beta_2+\frac{\beta_1-\beta_2}{1+\exp{(-x+\beta_3/|\beta_4|})}.
\end{equation}

\begin{table}[!t]
\caption{Effects of temporal sampling rate (stride) on FAVER when testing on the LIVE-YT-HFR database.}
\label{table:sampling-rate}
\setlength{\tabcolsep}{4pt}
\centering
\scriptsize
\begin{tabular}{l|cccc}
\toprule
\textbf{Model}                               & \begin{tabular}[c]{@{}c@{}}\textbf{Sampling} \\ \textbf{Stride} \end{tabular}& \textbf{SROCC↑}                 & \textbf{PLCC↑}           & \textbf{RMSE↓}           \\ \midrule
FAVER-Haar                   & 1 second    & 0.5864                                        & 0.6546                      & 9.1789\\ 
                    &
                   16 frames &\textbf{0.5973}                  & \textbf{0.6690} &          \textbf{9.0604}          \\ 
                          &          8 frames & 0.5737                       & 0.6413           & 9.3200          \\ 
                          &
                 4 frame  & 0.5880                 & 0.6491          & 9.2020          \\ \hline
FAVER-Db2                    & 1 second    & 0.6195                      & 0.6769             & 8.9525             \\
&
                 16 frames & 0.6219                  & 0.6784           & 8.8959          \\ 
                 &
                     8 frames & 0.6135                 & 0.6687          & 9.0220          \\ 
                     &
                 4 frames  & \textbf{0.6262}  & \textbf{0.6808} & \textbf{8.8470} \\ \hline
FAVER-Bior22                 & 1 second   & \textbf{0.6350}                                          & \textbf{0.6923}                      & \textbf{8.7964}                      \\  
                                    & 16 frames & 0.6262                   & 0.6853          & 8.8456          \\ 
                                    & 8 frames & 0.6136                   & 0.6737          & 8.9378          \\ 
                                    & 4 frames  & 0.6139                   & 0.6829          & 8.9177          \\ 
\bottomrule
\end{tabular}
\begin{tablenotes}
    \item[1] Sampling stride 1 second: sampled once per second (videos of any framerate) when extracting spatial and temporal features.
    \item[2] Sampling stride 16 (or 8, or 4) frames: the feature extracting module was executed every 16 (or 8, or 4) frames.
    \end{tablenotes}
\end{table}

\begin{table}[!t]
\centering
    \caption{Ablation Study of FAVER on LIVE-YT-HFR. The first five rows show the results for the reduced FAVER model using features only from Y, YUV, YGM, YLOG and YGMLOG. The FAVER-Spt, FAVER-Tmp-Db2, and FAVER-All-Db2 denote the spatial features, temporal features (using Db2), and the full FAVER model, respectively.}
    \label{table:ablation-study}
    \begin{tabular}{lccc}
    \toprule
    \textbf{Model}                 & \textbf{SROCC↑}  & \textbf{PLCC↑}  & \textbf{RMSE↓}   \\ \midrule
    Y                  & 0.4324 & 0.4991 & 10.5369 \\ 
    YUV                & 0.5275  & 0.6042 & 9.7655  \\ 
    YGM                & 0.3721  & 0.4785 & 10.7230 \\ 
    YLOG               & 0.4210  & 0.5027 & 10.4885 \\ 
    YGMLOG             & 0.3777 & 0.4625 & 10.8500   \\ 
    FAVER-Spt          & 0.5248 & 0.6035 & 9.8278  \\ 
    FAVER-Tmp-Db2      & {0.4540}                         & {0.5329}            & {10.2687}           \\
    FAVER-All-Db2  & \textbf{0.6195}                      & \textbf{0.6769}             & \textbf{8.9525}             \\ 
    \bottomrule
    \end{tabular}
\end{table}

\begin{table}[!t] 
\centering
\caption{Performance Comparison of all compared models on BVI-HFR.}
\label{table:bvi-hfr}
\begin{tabular}{lccc}
\toprule
\textbf{Model}                 & \textbf{SROCC↑}                  & \textbf{PLCC↑}           & \textbf{RMSE↓}            \\ \midrule
NIQE\cite{6353522}                  & 0.2247                 & 0.4194          & 16.8898          \\ 
BRISQUE\cite{mittal2012no}               & 0.2600                   & 0.4448          & 16.6404          \\ 
GM-LOG\cite{xue2014blind}                & 0.2778                    & 0.4647          & 16.3202          \\ 
VIDEVAL~\cite{9405420}               & 0.3449                  & 0.4742          & 16.2954          \\ 
HIGRADE\cite{kundu2017no}              & 0.2546                  & 0.4180           & 16.8538          \\ 
FRIQUEE\cite{FRIQUEE}                & 0.2387                  & 0.3733          & 16.8940          \\ 
RAPIQUE\cite{DBLP}               & 0.3037                 & 0.4631          & 16.3567          \\ 
TLVQM\cite{korhonen2019two}                  & 0.3734                  & 0.4908          & 16.1876           \\
CORNIA\cite{6247789}                & 0.2638                 & 0.4388          & 16.2733          \\ 
\midrule
FAVER-Haar           & \textbf{0.4117}                & \textbf{0.5212}                & \textbf{15.7612}         \\
FAVER-Db2         & \textbf{0.5008}                  & \textbf{0.5872}                & \textbf{14.8300} \\ 
FAVER-Bior22           & \textbf{0.5560}                   & \textbf{0.6390}         & \textbf{14.1079}  \\
\bottomrule
\end{tabular}
\end{table}

\subsection{Performance Comparison on LIVE-YT-HFR}
\label{ssec:live-yt-hfr}

We report the performance of all of compared models on the LIVE-YT-HFR dataset, in Table~\ref{table:performance-live-yt-hfr}. Three different versions of the entire (using all features) FAVER model using Haar, Db2, and Bior22 filters, were tested, and respectively denoted FAVER-Haar, FAVER-Db2 and FAVER-Bior22. It may be observed from Table~\ref{table:performance-live-yt-hfr} that the family of FAVER models significantly outperformed
the other compared blind VQA models by a large margin, with FAVER-Db2 and FAVER-Bior22 delivering the best performance. Fig.~\ref{fig:boxplots} shows the spreads of the
SROCC values for each evaluated NR-VQA model over the 100 iterations, showing that FAVER-bior22 had a much tighter confidence
interval than the other models, highlighting the robustness of
the algorithm. The scatter plots and best logistic curve fits of the model prediction scores against MOS using 5-fold cross-validation on LIVE-YT-HFR are shown in Fig.~\ref{fig:scatter}. It may be visually observed that, as compared with the other models, the predictions produced by FAVER were more consistent than the other compared models, with a smaller prediction variance. 

\subsection{Effects of Temporal Sampling Rate}
\label{ssec:sampling}

Video signals are highly redundant along the temporal dimension, especially along optical flow trajectories, providing significant opportunity for video compression~\cite{Lee:21}. Temporal distortions like stutter, judder, and motion blur affect these correlations. By appropriately modeling temporal natural video statistics, we can define temporal quality-aware features. A tradeoff exists between temporal feature density and model performance. To explore this issue, we varied the temporal sampling rate (or stride) of feature extraction in FAVER.

\begin{figure}[!t]
\centering
\def\xheight{0.76}
\footnotesize
\subfigure[SROCC]{\includegraphics[width=\xheight\linewidth]{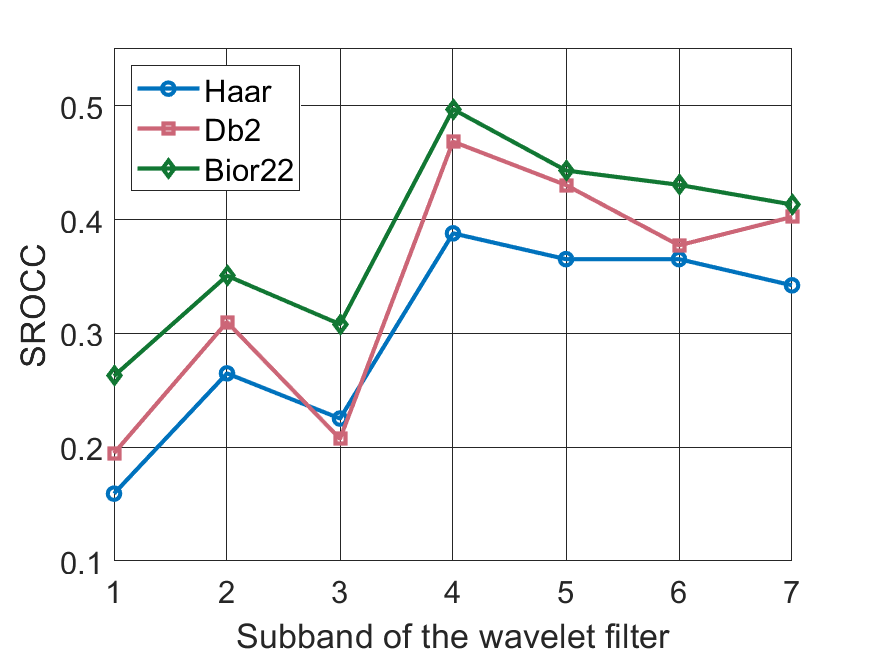}}
\subfigure[PLCC]{\includegraphics[width=\xheight\linewidth]{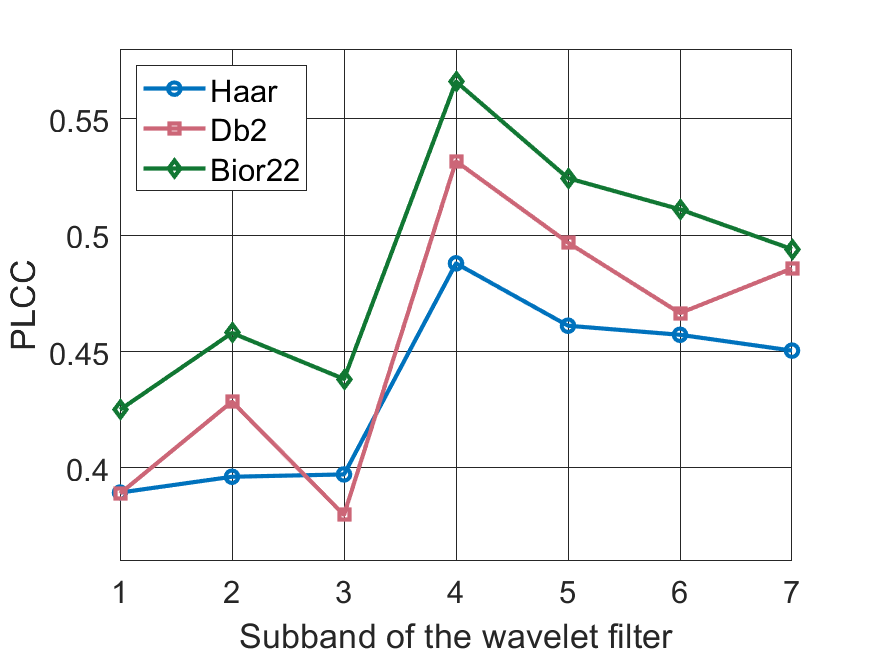}}
\caption{Contributions of subbands contained in the temporal responses in FAVER for each type of bandpass filter on LIVE-YT-HFR. The figure on the left shows performance as measured by SROCC, while the one on the right shows PLCC.}
\label{fig:hvs}
\end{figure}

\begin{table*}[!t]
\centering
\footnotesize
\caption{Complexity analysis of evaluated BVQA models on LIVE-YT-HFR. The time cost (seconds) is measured by averaging the seconds used for feature extraction on ten 1080p videos.}
\label{table:complexity}
\setlength{\tabcolsep}{10pt}
\begin{tabular}{l|r|rrr|c}
\toprule
 & Feat
  & \multicolumn{3}{c|}{Time (second/video@1080p)}  & Computaional complexity                                                                                           \\
                 Model      &     Dim                 & 30fps    & 60fps    & 120fps                                       \\ \midrule
BRISQUE                & 36                   & 3.07   & 3.13   & 3.07   & $\mathcal{O}(d^2NT)$                                                                                                                             \\
GM-LOG                  & 40                   & 2.54   & 2.46   & 2.46   & $\mathcal{O}((h+p)NT)$                                                                                                  \\
HIGRADE                & 216                  & 22.27  & 21.94  & 22.81  & $\mathcal{O}(3(2d^2+q)NT)$                                                                                                \\
FRIQUEE                & 560                  & 1063.80   & 1078.10   & 1070.20   & $\mathcal{O}(fd^2N+4N(log(N)+n^2))T)$                                                                                  \\
HOSA                   & 14.7k                & 2.11   & 1.96   & 1.95   & $\mathcal{O}(d^2K_1NT)$                                                                                                            \\
CORNIA                 & 10k                  & 20.47  & 20.22   & 20.75  & $\mathcal{O}(d^2K_1NT)$                                                                                                            \\
TLVQM                  & 75                   & 278.43 & 396.95 & 684.10 & $\mathcal{O}(h_1^2N+m^2K_2)T_1+(log(N)+h_2^2)NT_2))$      \\
VIDEVAL                & 60                   & 518.80 & 984.34 & 2045.80   & $\mathcal{O}(fh_1^2N+m^2K_3)T_1+h_2^2NT_2)$           \\
RAPIQUE                & 3884                 & 38.34  & 37.42  & 38.09  & $\mathcal{O}(((4d^2+h)N+fd^2N+(8+7d^2)N)T_2)$                          \\
\midrule
FAVER-Haar            & 748                  & 29.86  & 33.13   & 32.13    & $\mathcal{O}(((5d^2+h)N+(8+7d^2)N)T_2)$                               \\
FAVER-Db2            & 748                  & 48.01  & 47.79    & 47.60  & $\mathcal{O}(((5d^2+h)N+(23+7d^2)N)T_2)$                             \\
FAVER-Bior22            & 748                  & 56.51  & 56.63 & 56.23  & $\mathcal{O}(((5d^2+h)N+(31+7d^2)N)T_2)$ \\ \bottomrule                            
\end{tabular}
\begin{tablenotes}
        \item[1] Feat Dim: dimension of features; $d$: window size; $N$: number of pixels per frame; $T$:number of frames computed for feature extraction; $h$: filter size; $q$: gradient kernel size; $K_1$: codebook size; $m$: motion estimation block size; $K_2$: number of key; $n$: neighborhood size in DNT; $p$: probility matrix size; $f$: number of color spaces; $K_3$: number of key points;  $T_1$: total number of frames divided by 2; $T_2$: number of frames sampled at 1 fr/sec. 
\end{tablenotes}
\end{table*}

In Table~\ref{table:sampling-rate}, we report the performance comparison between FAVER models using different sampling rates of (strides) temporal feature extraction. In our implementation, we sought to maximize compute efficiency without sacrificing performance. Specifically, the spatial features in FAVER are computed on only one frame each second; this effective subsampling achieves significant speedup with little loss of performance. Likewise, four different sampling strides on the temporal features among 1 second, 16 frames, 8 frames, and 4 frames, are compared for each temporal filter used. Note that each temporal filter takes different number of frames as inputs no matter what stride is used. Specifically, 1) when using Haar filters, temporal features are computed on 8 consecutive frames within each one-second interval; 2) when using Db2 filters, the temporal features are computed on 23 frames each second; 3) when using Bior22, the temporal features are computed on 31 frames each second. It may be observed from Table~\ref{table:sampling-rate} that this very significant gain in efficiency causes only a slight reduction in prediction power when using a stride of 1 second against denser sampling rates. Thus, our final FAVER model family uses this sampling strategy. To be clear, the number of frames utilized corresponds to the filter length, and in each case, one temporal response (for each feature) is computed each second.

\subsection{Feature Importance Analysis}
\label{ssec:feature-importance}

We also conducted an ablation study to analyze the contributions of the individual features of FAVER. In this experiment we trained the model on a subset of the feature maps (e.g., $YUV$ means that only features extracted from the $YUV$ maps were used). As shown in Table~\ref{table:ablation-study}, adding features from the gradient and LoG luminance domains significantly increased performance. We also observed that appending features from the chroma channel ($U, V$) greatly boosted performance on LIVE-YT-HFR. 

From Table~\ref{table:ablation-study} it may also be observed that applying only spatial features (FAVER-Spt) or only temporal features (FAVER-Tmp-Db2) yielded fair performance, but combining them (FAVER-All-Db2) delivered significantly better performance. This affirms the notion that the spatial and temporal features capture complementary perceptual quality information on VFR videos.

\subsection{Contribution of Temporal Subbands}

The 3-level bandpass filters (Haar, Db2, Bior22) used in the three FAVER prototypes are each expressed in seven subbands (ignoring the lowest subband). Next, we investigated the relative importances of the subbands by training and testing reduced FAVER models consisting of each single subband on LIVE-YT-HFR testbed. As Fig.~\ref{fig:hvs} shows, for all three types of bandpass filters, the middle-frequency bands delivered higher correlations against human judgments in terms of both SROCC and PLCC, for all filter banks. This behavior is consistent with the shape of the temporal contrast sensitivity function (CSF)~\cite{robson1966spatial}.
    
\subsection{Performance Comparison on BVI-HFR}
\label{ssec:bvi-hfr-live-vqc}

BVI-HFR~\cite{8531714} is the only other sizeable VFR-VQA database. It consisting of 22 source sequences spanning 4 frame rates: 15, 30, 60, and 120 fps. Unlike LIVE-YT-HFR, it contains videos that were subsampled in time by spatially averaging frames. In this manner of subsampling, motion blur is introduced but stutter less so. BVI-HFR also does not include varied compression distortions. We again randomly divided the dataset into 80$\%$  and 20$\%$ subsets for training and test, ensuring no overlapping of contents. Since this dataset contains a limited number of videos having largely insignificant variation across framerates, we found it difficult for no-reference models to be fully trained, yielding relatively low correlations as compared to full-reference models~\cite{madhusudana2021subjective}. Yet, as summarized in Table~\ref{table:bvi-hfr}, FAVER outperformed all the competing BVQA models.


\subsection{Complexity Analysis}
\label{ssec:complexity}

We also carried out a computational complexity analysis on the evaluated models. For a fair comparison, all the experiments were conducted on the same computer, a XiaoXinPro-13ARE, with an AMD Ryzen 7 4800U with Radeon Graphics@1.80GHz, and 16G RAM. The models were implemented using their original releases on MATLAB R2020b. We also accounted for the video framerates, and recorded the time costs at 30fps, 60 fps, and 120 fps. Table~\ref{table:complexity} tabulates the feature dimension, actual time cost against framerate, as well as the theoretical complexity of the compared BVQA models. As it may be seen, the three FAVER variants are efficient as compared to other top-performing BVQA models like TLVQM and VIDEVAL. FAVER is \textbf{6x-9x} faster than TLVQM, and \textbf{9x-17x} faster than VIDEVAL on 1080p@30fps videos. It is also worth mentioning that FAVER maintains a similar computational cost as the video framerate is increased, while those of TLVQM and VIDEVAL scale linearly with respect to framerate. Moreover, FAVER obtained comparable complexity as the most efficient model, RAPIQUE, but it utilizes far fewer features while delivering significantly better performance on VFR videos.

\section{Conclusion}
\label{sec:conclusion}

We proposed the first robust, effective and efficient BVQA model focused on perceptual quality assessment of VFR videos, dubbed FAVER. FAVER utilizes the temporal natural video statistics of bandpass filtered videos to capture and represent aspects of temporal video quality. FAVER is trained using an ensemble of learners to temporal and spatial quality scores which are then combined into a final quality score. Our extensive experimental results show that FAVER delivers accurate and stable predictions of video quality on HFR/VFR database against human judgments, exceeding the performances of previous video quality models. FAVER is computationally efficient, with nearly constant complexity as the video framerate is increased. We believe this work will facilitate and inspire future research efforts on the quality assessment and intelligent compression of VFR videos.


%




\ifCLASSOPTIONcaptionsoff
  \newpage
\fi



%


\bibliographystyle{IEEEtran}
\bibliography{IEEEfull}

%








\end{document}